\documentclass[draftclsnofoot,onecolumn]{IEEEtran}

\normalsize

\ifCLASSINFOpdf
\else
\fi

\usepackage{graphicx}
\usepackage{bm}
\usepackage{amsfonts}
\usepackage{amsmath}
\usepackage{amssymb}
\usepackage{times}
\usepackage{subfigure}
\usepackage{latexsym,bm,amsmath,amssymb}
\usepackage{cite}
\usepackage{hhline}

\usepackage{multirow} 
\usepackage{amsmath}
\usepackage{xcolor}
\usepackage{epstopdf}
\usepackage{algorithmicx,algorithm}

\begin{document}

\title{Fundamentals of Delay-Doppler Communications: Practical Implementation and Extensions to OTFS}

\author{Shuangyang Li,~\IEEEmembership{Member,~IEEE,}
        Peter Jung,~\IEEEmembership{Member,~IEEE,}
        Weijie Yuan,~\IEEEmembership{Member,~IEEE,}
        Zhiqiang Wei,~\IEEEmembership{Member,~IEEE,}
        Jinhong Yuan,~\IEEEmembership{Fellow,~IEEE,}
        Baoming Bai,~\IEEEmembership{Senior Member,~IEEE,}
        and~Giuseppe Caire,~\IEEEmembership{Fellow,~IEEE}
\thanks{This paper was presented in part at IEEE Global Communication Conference 2023~\cite{Shuangyang2023Globecom}.}
\thanks{S.~Li, P.~Jung, and G.~Caire are with the Faculty of Electrical Engineering and Computer Science, Technical University of Berlin, Berlin, 10587, Germany (e-mail: shuangyang.li, pter.jung, caire@tu-berlin.de).}
\thanks{
W.~Yuan is with the Department of Electrical and Electronic Engineering,
Southern University of Science and Technology, Shenzhen 518055, China
(e-mail: yuanwj@sustech.edu.cn).
}
\thanks{Z.~Wei is with the School of Mathematics and Statistics, Xi'an Jiaotong University, Xi'an 710049, China (e-mail: zhiqiang.wei@xjtu.edu.cn).}
\thanks{J. Yuan is with the School of Electrical Engineering and Telecommunications, University of New South Wales, Sydney, NSW 2052, Australia (e-mail: j.yuan@unsw.edu.au).}
\thanks{B.~Bai is with the State Key Lab.~of ISN, Xidian University, Xi'an 710071, China.
(e-mail:~bmbai@mail.xidian.edu.cn).}
}

\maketitle

\begin{abstract}
The recently proposed orthogonal time frequency space (OTFS) modulation, which is a typical Delay-Doppler (DD) communication scheme, has attracted significant attention thanks to its appealing performance over doubly-selective channels.
In this paper, we present the fundamentals of general DD communications from the viewpoint of the Zak transform. We start our study by constructing DD domain basis functions aligning with the time-frequency (TF)- consistency condition, which are globally quasi-periodic and locally twisted-shifted. We unveil that these features are translated to unique signal structures in both time and frequency, which are beneficial for communication purposes. Then, we focus on the practical implementations of DD Nyquist communications, where we show that rectangular windows achieve perfect DD orthogonality, while truncated periodic signals can obtain sufficient DD orthogonality. Particularly, smoothed rectangular window with excess bandwidth can result in a slightly worse orthogonality but better pulse localization in the DD domain. Furthermore, we present a practical pulse shaping framework for general DD communications and derive the corresponding input-output relation under various shaping pulses. Our numerical results agree with our derivations and also demonstrate advantages of DD communications over conventional orthogonal frequency-division multiplexing (OFDM).
\end{abstract}


\IEEEpeerreviewmaketitle

\section{Introduction}
Next generation wireless networks are expected to provide high throughput and ultra-reliable communications services to facilitate the various emerging applications. To meet the stringent requirements imposed by these emerging applications, such as high frequency bands, high mobility, conventional wireless waveforms, e.g., orthogonal frequency-division multiplexing (OFDM), require sophisticated adaptation, which not only complicates the system design but also degrade the system performance potentially.
In light of this, new wireless waveforms need to be developed.

Orthogonal time frequency space (OTFS) modulation was recently proposed in~\cite{Hadani2017orthogonal} as a means for next generation wireless communications. The advantages of OTFS come from the symbol placement in the delay-Doppler (DD) domain, which allows the information symbol to have a direct interaction to the DD domain channels. As a result, appealing DD domain channel properties are naturally exploited, including quasi-static, path separability, and compactness~\cite{Zhiqiang_magzine,hlawatsch2011wireless}. This channel exploitation has translated into various improvements of communication performance, as evidenced by many existing works~\cite{Yuan2023survey,ShuangyangLetterPartI,ShuangyangLetterPartII,ShuangyangLetterPartIII}. In~\cite{Raviteja2019embedded}, a DD domain channel estimation scheme based on the embedded pilot was proposed, where one strong pilot symbol is placed in the DD domain with a sufficiently large guard space. Thanks to the DD domain path separability, good channel estimation performance can be achieved by simply comparing the received and the transmitted pilot symbols. The error performance of OTFS was studied in~\cite{Raviteja2019effective}, where the authors have shown that OTFS almost surely achieves the full channel diversity over sparse Rayleigh fading channels even with a relatively small frame size. This conclusion is further extended in~\cite{Li2020performance}, which reported that coded OTFS systems have an important diversity and coding gain tradeoff depending on the number of resolvable paths in the channel. Consequently, OTFS only requires a relaxed code design comparing to OFDM. Furthermore, the DD domain path separability gives rise to novel MIMO communication designs using OTFS waveforms. Particularly, path-oriented precoding designs have shown good compatibility to both point-to-point (P2P) MIMO and multi-user (MU) MIMO transmissions in terms of both achievable rates and computational complexity~\cite{Mengmeng2023P2PMIMO,LSY_THP}.
In addition, recent works on integrated communications and sensing (ISAC) have also demonstrated the great potential of OTFS~\cite{Dehk_TWC,Shuangyang_ISAC}.

Let us have a historical recap of the OTFS literature. In the early papers~\cite{Hadani2017orthogonal,Raviteja2018interference,RezazadehReyhani2018analysis}, OTFS was implemented based on the overlay of OFDM by using the two-dimensional (2D) inverse symplectic finite Fourier transforms (ISFFT), which is known as the two-stage implementation. Such an implementation demonstrates a strong compatibility with the main stream OFDM standards but did not highlight the unique properties of OTFS promised by the Zak transform, which is the fundamental mathematical tool connecting the DD domain and the time/frequency domain. After several years of prosperity, OTFS has gradually become a popular topic in both academia and industry. However, most of the OTFS studies were still focused on the two-stage implementation of OTFS until now. Noticeably, the two-stage implementation is limited by the OFDM structure, where a time-frequency (TF) domain pulse, commonly a rectangular window with duration of an OFDM symbol, was adopted for carrying the precoded (using ISFFT) information symbols. Note that the ISFFT spreads each DD domain information symbol to all TF domain symbols, which are carried by different TF domain pulses. Consequently, each DD domain information symbol is modulated onto many different pulses, resulting in the so-called ``pulse discontinuity''~\cite{Lin2022ODDM}. The pulse discontinuity may introduce abrupt changes of the transmitted signal, which can cause a large out-of-band (OOB) emission and thereby degrading the system performance.
However, the impact of pulse discontinuity is almost invisible in numerical simulations if the OOB issue was not considered, which is often the case when the system was
evaluated by Monte Carlo methods developed based on the discrete channel model, e.g.,~\cite{Raviteja2019practical}.
To solve the OOB issue, many OTFS variants have been proposed recently. The orthogonal delay-Doppler division multiplexing (ODDM) modulation was proposed in~\cite{Lin2022ODDM}, where a realizable orthogonal basis with respect to DD resolutions was constructed using the staggered multi-tone modulation.
In this context, the authors further proposed delay-Doppler multicarrier (DDMC) modulation using delay-Doppler orthogonal pulses (DDOP) and demonstrated that sufficient DD orthogonality can be achieved by using periodically extended root-raised cosine (RRC) pulses or a Nyquist pulse train~\cite{Hai2022DDOP}.
Furthermore, a framework of pulse shaping on DD plane was reported in~\cite{bayat2023unified}, where a low-complexity pulse shaping structure based on fast convolution was proposed. The proposed framework is compatible to various shaping pulses, and an end-to-end discrete system model was also provided.

More recently, researchers have realized the importance of the Zak transform to OTFS. The Zak transform was originally introduced in the field of
solid state physics by J. Zak~\cite{janssen1988zak} and was then extended to the field of signal processing~\cite{janssen1988zak}. It is a mathematical tool that highlights the physical interpretation between time/frequency and DD.
Compared to the OTFS implementation based on OFDM transceivers, implementation using the Zak transform requires less complexity and preserves clear physical insights. A main feature of OTFS transmissions based on the Zak transform is that it generally does not require the overlay with OFDM. Therefore, it is also known as the one-stage implementation. The discrete Zak transform (DZT)-based OTFS realization was proposed in~\cite{Lampel2022ICC}, where input-output relation of OTFS was studied and discussed from the DZT viewpoint. However, due to the discrete nature of DZT, the DZT-based OTFS exhibits a degraded performance when the DD resolutions are not sufficient (commonly known as the ``fractional delay and Doppler''). In~\cite{mohammed2021derivation}, a Zak transform based implementation of OTFS was presented. Specifically, the DD domain continuous symbol carrier (known as the basis function) was derived according to the Zak transform, which was shown to be sufficiently localized. Furthermore, bandwidth-limited and time-limited window functions were applied to truncate the basis functions, where the author showed that rectangular windows can achieve the perfect DD orthogonality.
This work has been extended to the OTFS 2.0 modulation recently in~\cite{MohammedBITSpart1,Mohammed2023part2}. The OTFS 2.0 highlights the DD domain information transmission based on the Zak transform, where the end-to-end input-output relation was characterized by the \emph{twisted-convolution} in the DD domain. In particular, the OTFS 2.0 intentionally precoded information symbols to satisfy the quasi-periodicity property of the Zak transform, and apply DD domain shaping pulses at both the transmitter and receiver to convey information. As a result, OTFS 2.0 exhibits a mathematically simple input-output relation defined purely in the DD domain. However, all real-world signal transmissions are essentially implemented in time, and unfortunately, not all
DD domain signals are realizable in time\footnote{As will be shown later, only DD domain signals satisfy the quasi-periodicity are realizable in time.}. Therefore, how to implement OTFS and in general DD communications in practice require further studies.

In this paper, we aim to fill the gap between the DD communication theory based on the Zak transform and its \textbf{practical} implementation.
Specifically, we present the fundamentals of practical DD communications based on the Zak transform without relying on the overlay of the transceivers of TF domain multicarrier waveforms, e.g., OFDM. In contrast of the previous works~\cite{MohammedBITSpart1,Mohammed2023part2}, we highlight the practical implementation of DD Nyquist communications using realizable pulse shaping filters, where an end-to-end input-output relation is also derived from the communication theory viewpoint.
Our proposed DD communication framework enjoys sufficient DD orthogonality, which may be the exact motivation of OTFS.
More importantly, we highlight the physical interpretation of DD and time/frequency from a signal processing point of view. We show that periodicities in time and frequency will result in localizations in Doppler and delay, according to the theory of Zak transform. However, exact periodicity requires infinite time and frequency resources, which are not realizable in practice. Therefore, truncating periodic time and frequency signals are well motivated, and the DD localizations are degraded to DD orthogonality due to the truncation.
The main contributions of this paper are summarized as follows.
\begin{itemize}
  \item We define a group of equally-spaced basis functions corresponding to information symbols, which are constructed in a special way such that their transformations in both time and frequency are consistent.
      Such basis functions naturally incorporate the twisted-convolution insight in the DD domain, which exhibit quasi-periodicity globally while are twisted-shifted locally. Particularly, the constructed basis function allows straightforward calculation of its ambiguity function, which is in line with the DD domain pulse shaping and matched-filtering. Furthermore, we unveil that the DD domain global feature of such functions translates into a train of pulses, while their local feature translates into signal tones in both time and frequency. This unique structure is known as the ``pulsone'' in the literature of OTFS.
  \item We further present the practical realization of basis function by applying practical filters to the ideal basis functions for symbol transmissions. We introduce the TF-consistent pulse shaping for deriving the practical basis functions, where both time domain and frequency domain window functions are applied for obtaining a roughly time-limited and bandwidth-limited basis function. We derive the corresponding DD domain representation and ambiguity function of such basis functions, and further demonstrate that basis functions enjoying sufficient localization and orthogonality can be achieved by applying truncated periodic signals for windowing. More importantly, we verify that the rectangular windows enjoy perfect DD orthogonality.
   \item We present the practical pulse shaping implementation for DD communications in the time domain by introducing reasonable approximations. Based on the proposed implementation, we further derive the input-output relation of general DD communications with various shaping pulses over underspread channels. Particularly, we demonstrate that the above input-output relation are well-characterized by the ambiguity function of the basis function in the asymptotical regime, which yields essentially the same results as in~\cite{MohammedBITSpart1,Mohammed2023part2}. Furthermore, we also verify that the above input-output relation using rectangular windows converges to that of OTFS implemented using OFDM transceivers with rectangular TF pulses~\cite{Raviteja2019practical}.
   \item We provide numerical results of DD communications using various shaping pulses in terms of the bit error rate (BER), pragmatic capacity~\cite{Lorenzo2022fair,Kavcic2003binary}, and power spectral density (PSD). The practical advantages of the proposed scheme are verified based on these results.

\end{itemize}

\emph{Notations:} The blackboard bold letters ${\mathbb{A}}$, ${\mathbb{Z}}$, and ${\mathbb{E}}$ denote the energy-normalized constellation set, the integer number field, and the expectation operator, respectively;
``$*$" and ``$ \otimes $'' denote the convolution and kronecker product, respectively; ``${{\left[ \cdot \right]}_M}$'' denotes the module-$M$ operation.
``${\left( {\cdot} \right)^*}$'' denotes the conjecture operation.  ${\rm sinc}\left( x \right)$ is the sinc function{\footnote{As commonly defined, we have ${\rm{sinc}}\left( 0\right)=0$.}} defined by ${\rm{sinc}}\left( x \right) \buildrel \Delta \over = \frac{{\sin \left( \pi x \right)}}{\pi x}$.

We will interchangeably use two sets of representations to describe the same signal in different domains. For a time domain signal $x(t)$, its Fourier transform and its Zak transform are denoted by $X(f)$ and ${{\cal Z}_x}\left( {\tau ,\nu } \right)$, respectively. This type of representations highlights the transformation among different domains.
Equivalently, we also use subscripts to highlight the domain in which signal is defined, e.g., ${\Phi}_{\rm T}(t)$, ${\Phi}_{\rm F}(f)$, and ${\Phi}_{\rm DD}(\tau, \nu)$ denote the same signal represented in time, frequency, and DD domains, respectively.

\section{Preliminaries on Zak Transform and DD Domain Pulses}
In this section, we will review some fundamental properties of the Zak transform that are available in the literature~\cite{janssen1988zak,Bolcskei1994Gabor}. We note that the Zak transform plays an important role in the context of the Gabor expansion~\cite{Bolcskei1994Gabor}, and its significance has been widely explored in many engineering aspects including image processing~\cite{Teuner1993adaptive}, texture segmentation~\cite{Dunn1995optimal}, and more recently, wireless communications~\cite{Hadani2017orthogonal}. Note that the Zak transform is a version of the Poisson summation formula~\cite{grochenig2013foundations} and therefore it holds for general Schwartz functions. In fact, it is well-defined almost everywhere in the ${\cal L}^2$ space, see Lemma 8.2.1 in~\cite{grochenig2013foundations}.
Specifically, the Zak transform can be defined equivalently for both time domain signals and frequency domain signals as shown in the following~\cite{grochenig2013foundations}, where we assume that the underlying time domain signal and its Fourier transform are well-defined in the Wiener space.

\textbf{Definition~1} (\emph{The Zak Transform}):
Let $x\left( t \right)$ be a complex-valued time-continuous function, whose Fourier transform is given by $X\left( f \right)$. Furthermore, let $T$ be a positive constant. Then, the Zak transform can be defined equivalently in both time and frequency by~\cite{janssen1988zak,Bolcskei1994Gabor}
\begin{align}
{{\cal Z}_x}\left( {\tau ,\nu } \right)=\left( {\cal ZT}_{\rm T} \; x  \right)\left(  {\tau ,\nu } \right)\buildrel \Delta \over = \sqrt T \sum\limits_{k =  - \infty }^\infty  {x\left( {\tau  + kT} \right){e^{ - j2\pi k\nu T}}}
\label{Zak_Transform_def}
\end{align}
and
\begin{align}
{{\cal Z}_x}\left( {\tau ,\nu } \right)= \left( {\cal ZT}_{\rm F} \; X  \right)\left(  {\tau ,\nu } \right)\buildrel \Delta \over = \frac{1}{{\sqrt T }}{e^{j2\pi \nu \tau }}\sum\limits_{k =  - \infty }^\infty  {X\left( {\nu  + \frac{k}{T}} \right){e^{j2\pi k\frac{\tau }{T}}}} , \label{ZT_def_FT}
\end{align}
respectively, for $ -\infty   < \tau  < \infty $ and $ -\infty  < \nu  < \infty$. Here,
 ${\cal ZT}_{\rm T}$ and ${\cal ZT}_{\rm F}$ are linear mappings that map signals in time or frequency to DD.

We highlight that the convergence in~\eqref{Zak_Transform_def} and~\eqref{ZT_def_FT} holds almost everywhere in the ${\cal L}^2$ space~\cite{grochenig2013foundations}. In what follows, we restrict ourselves by only considering cases where the above convergence holds without explicitly mentioning.
Conversely, the inverse Zak transform gives the corresponding time domain and frequency domain signals based on the DD domain signal response, and it is defined in the following.

\textbf{Definition~2} (\emph{Inverse Zak Transform}):
Given a signal $x\left( t \right)$, whose Fourier transform and Zak transform are  given by $X\left( f \right)$ and ${{\cal Z}_x}\left( {\tau ,\nu } \right)$, respectively, we have~\cite{janssen1988zak,Bolcskei1994Gabor}
\begin{align}
x\left( t \right) = \left( {\cal IZT}_{\rm T} \; {{\cal Z}_x}\right) \left( t \right) \buildrel \Delta \over = {\sqrt{T}}\int_0^{\frac{1}{T}} {{{\cal Z}_x}\left( {t ,\nu } \right)} {\rm{d}}\nu  \label{Inverse_ZT}.
\end{align}
and
\begin{align}
X\left( f \right) = \left( {\cal IZT}_{\rm F} \; {{\cal Z}_x}\right) \left( f \right)\buildrel \Delta \over =\frac{1}{{\sqrt T }}\int_0^T {{{\cal Z}_x}\left( {\tau ,f } \right){e^{ - j2\pi f \tau }}} {\rm{d}}\tau  \label{ZT_to_FT}.
\end{align}
Here, ${\cal IZT}_{\rm T}$ and ${\cal IZT}_{\rm F}$ are linear mappings that map signals in DD to time or frequency.

\begin{table}[t]
\centering
\caption{Important Properties of the Zak Transform}
\begin{tabular}{|c|c|c|c|}
\hline
                                          & \begin{tabular}[c]{@{}c@{}}Signal in  time\end{tabular} & \begin{tabular}[c]{@{}c@{}}Signal in  frequency\end{tabular} & \begin{tabular}[c]{@{}c@{}}After Zak transform\end{tabular}
                                                                                                     \\ \hline
Delay-Doppler shifting                    & ${e^{j2\pi {\nu _1}\left({t-{\tau _1}}\right)}} x\left( {t - {\tau _1}} \right)$                                                        & $ {e^{-j2\pi f {\tau _1}}} X\left( {f - {\nu _1}} \right)$                                                              & $ {e^{j2\pi {\nu _1}\left( {\tau  - {\tau _1}} \right)}}{{\cal Z}_x}\left( {\tau  - {\tau _1},\nu  - {\nu _1}} \right)$                                                       \\ \hline
Multiplication in time                    & $x\left( t \right)y\left( t \right) $                                                        & $X\left( f \right)*Y\left( f \right) $                                                            & $ {\sqrt T}\int_0^{\frac{1}{T}} {{{\cal Z}_x}\left( {\tau ,\nu '} \right){{\cal Z}_y}\left( {\tau ,\nu  - \nu '} \right)} {\rm d}\nu '$                                                         \\ \hline
Convolution in time & $x\left( t \right)*y\left( t \right)$                                                         & $X\left( f \right)Y\left( f \right) $                                                               & $\frac{1}{{\sqrt T }}\int_0^T {{{\cal Z}_x}\left( {\tau  - \tau ',\nu } \right){{\cal Z}_y}\left( {\tau ',\nu } \right)} {\rm d}\tau '$                                                            \\ \hline
\end{tabular}
\label{Zak_properties}
\end{table}

In Table~\ref{Zak_properties}, we summarize some important properties of the Zak transform with respect to the time and frequency operations, which will be frequently used throughout this paper. In addition, the following two lemmas will also be widely used throughout the paper, whose proofs can be found in~\cite{janssen1988zak,Bolcskei1994Gabor}.

\textbf{Lemma~1} (\emph{Quasi-Periodicity}):
The Zak transform is quasi-periodic along the delay axis with period $T$ and periodic along the Doppler axis with period ${\frac{1}{T}}$, i.e.,
\begin{align}
{{\cal Z}_x}\left( {\tau  + T,\nu } \right) = {e^{j2\pi T\nu }}{{\cal Z}_x}\left( {\tau ,\nu } \right)\label{Zak_quasi_periodicity_delay},
\end{align}
and
\begin{align}
{{\cal Z}_x}\left( {\tau ,\nu  + \frac{1}{T}} \right) = {{\cal Z}_x}\left( {\tau ,\nu } \right)  \label{Zak_periodicity_Doppler}.
\end{align}

\textbf{Lemma~2} (\emph{Zak Transform vs. Ambiguity Function}):
The cross ambiguity function for functions $x\left( t \right)$ and $y\left( t \right)$ is defined by
\begin{align}
{A_{x,y}}\left( {\tau ,\nu } \right) \buildrel \Delta \over = \int_{ - \infty }^\infty  {x\left( t \right){y^*}\left( {t - \tau } \right){e^{ - j2\pi \nu \left( {t - \tau } \right)}}{\rm{d}}t}  = \int_{ - \infty }^\infty  {X\left( f \right){Y^*}\left( {f - \nu } \right){e^{j2\pi f\tau }}{\rm{d}}f} , \label{C4_Ambiguity_function}
\end{align}
where ${X\left( f \right)}$ and ${Y\left( f \right)}$ are the corresponding Fourier transforms of $x\left( t \right)$ and $y\left( t \right)$, respectively.
Then, given ${{\cal Z}_x}\left( {\tau ,\nu } \right)$ and ${{\cal Z}_y}\left( {\tau ,\nu } \right)$, the Zak transforms of $x\left( t \right)$ and $y\left( t \right)$, we have
\begin{align}
{{\cal Z}_x}\left( {\tau ,\nu } \right){\cal Z}_y^*\left( {\tau ,\nu } \right) = \sum\limits_{n =  - \infty }^\infty  {\sum\limits_{m =  - \infty }^\infty  {{A_{x,y}}\left( {nT,\frac{m}{T}} \right)} } {e^{ - j2\pi n\nu T}}{e^{j2\pi \frac{m}{T}\tau }} .\label{C4_ZT_AF}
\end{align}
Conversely, we have
\begin{align}
{A_{x,y}}\left( {nT,\frac{m}{T}} \right) = \int_0^T {\int_0^{\frac{1}{T}} {{{\cal Z}_x}\left( {\tau ,\nu } \right){\cal Z}_y^*\left( {\tau ,\nu } \right)} } {e^{ - j2\pi \frac{m}{T}\tau }}{e^{j2\pi n\nu T}} {\rm{d}}\nu {\rm{d}}\tau.\label{C4_AF_ZT}
\end{align}

Finally, we highlight that not all DD domain signals satisfy the properties of the Zak transform and applying the inverse Zak transform to arbitrary DD domain signals may not yield meaningful time or frequency signals.
Therefore, in this paper, we focus on the application of practical DD domain shaping pulses that satisfy~\eqref{Zak_quasi_periodicity_delay} and~\eqref{Zak_periodicity_Doppler} and therefore have meaningful time and frequency representations given by~\eqref{Inverse_ZT} and~\eqref{ZT_to_FT}.
For such pulses, the so-called ``fundamental rectangle'' is an important concept, which is a range of DD components of size $\tau  \in \left[ {0,T} \right)$ and $\nu  \in \left[ {0,\frac{1}{T}} \right)$. As suggested in Lemma~1, DD domain pulses satisfying the quasi-periodicity property must be sufficiently described by the signal behaviour in the fundamental rectangle~\cite{Bolcskei1994Gabor}. This fact will be used in the following part of this paper for constructing DD domain basis functions.

\section{Fundamentals of Delay-Doppler Domain Signaling}

\begin{figure}
\centering
\includegraphics[width=0.8\textwidth]{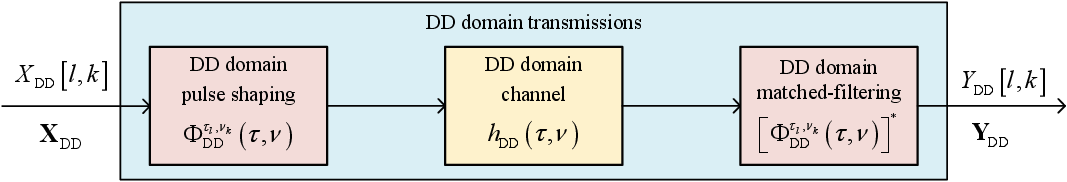}
\caption{The considered DD domain transmission for communications.}
\label{System_model}
\centering
\end{figure}
In this section, we focus on the application of Zak transform for communication systems, where the transmitted continuous signal is constructed by modulating a set of discrete information symbols using basis functions.
Specifically, we consider a type of DD domain signal transmissions for communication as shown in Fig.~\ref{System_model}, where
a set of discrete DD domain symbols are linearly modulated onto a family of continuous DD domain waveforms, i.e., \emph{DD domain basis functions}.
Let ${\bf X}_{\rm DD}$ be the DD domain symbol matrix of size $M \times N$, where $M$ and $N$ are numbers of delay and Doppler bins, respectively. Let ${X}_{\rm DD}\left[l,k\right]$ be the $\left(l,k\right)$-th element of ${\bf X}_{\rm DD}$, which is modulated onto the corresponding DD domain basis function ${\Phi _{{\rm{DD}}}^{{\tau _l},{\nu _k}}\left( {\tau ,\nu } \right)}$ via DD domain pulse shaping. Therefore, the considered linearly modulated DD communication signal is formulated as
\begin{align}
{s_{{\rm{DD}}}}\left( {\tau ,\nu } \right) = \sum\limits_{l = 0 }^{M-1}  {\sum\limits_{k = 0 }^{N-1}  {{X_{{\rm{DD}}}}\left[ {l,k} \right]\Phi _{{\rm{DD}}}^{{\tau _l},{\nu _k}}\left( {\tau ,\nu } \right)} } ,\label{DD_modulated_signal}
\end{align}
where ${\Phi _{{\rm{DD}}}^{{\tau _l},{\nu _k}}\left( {\tau ,\nu } \right)}$ is the DD domain basis function with offset $\tau_l$ and $\nu_k$ and it will be detailed in the coming subsection.

In this paper, we consider an arbitrary DD domain channel ${h_{{\rm{DD}}}}\left( {\tau ,\nu } \right)$, which is written as~\cite{hlawatsch2011wireless}
\begin{align}
{h_{{\rm{DD}}}}\left( {\tau ,\nu } \right) = \sum\limits_{p = 1}^P {{h_p}\delta \left( {\tau  - {\tilde \tau _p}} \right)} \delta \left( {\nu  - {\tilde \nu _p}} \right). \label{DD_channel}
\end{align}
In~\eqref{DD_channel}, $P$ can be interpreted as the number of resolvable paths, $h_p$, $\tilde\tau _p$, and $\tilde\nu _p$ are the channel coefficient, delay, and Doppler shift corresponding to the $p$-th path, respectively, where $\tilde\tau _p = \tilde\tau _{p'}$ and $\tilde\nu _p = \tilde\nu _{p'}$ for $p \ne p'$ do not hold simultaneously.
Furthermore, we consider that ${h_{{\rm{DD}}}}\left( {\tau ,\nu } \right)$ satisfy the \emph{crystallization} condition~\cite{MohammedBITSpart1,Mohammed2023part2}, which requires that
\begin{align}
{\tilde \tau _{\max }} - {\tilde \tau _{\min }} < T \quad  {\rm{  and   }}\quad {\tilde \nu _{\max }} - {\tilde \nu _{\min }} < \frac{1}{T}, \label{crystallization}
\end{align}
respectively. Here, ${\tilde \tau _{\max }}$ and ${\tilde \tau _{\min}}$ are the maximum and minimum values of the delays, while ${\tilde \nu _{\max }}$ and ${\tilde \nu _{\min}}$ are the maximum and minimum values of the Doppler shifts. Essentially, the crystallization condition suggests that the channel is \emph{underspread}~\cite{hlawatsch2011wireless,Peter2007WSSUS}, which can be achieved by carefully selecting $T$ with respect to the channel condition. Without loss of generality, we further assume that $\tilde \tau  \in \left[ {0,T} \right),\tilde \nu  \in \left[ { - \frac{1}{{2T}},\frac{1}{{2T}}} \right)$, which can be achieved by synchronizing the signal according to ${\tilde \tau _{\min }}$ and ${\tilde \nu _{\min}}$.

With the considered communication channel given by~\eqref{DD_channel}, the received DD domain signal $r_{{\rm{DD}}}\left( \tau, \nu \right)$ can be derived by the following theorem.

\textbf{Theorem~1} (\emph{DD Domain Twisted Convolution}):
The DD domain received signal ${r_{{\rm{DD}}}}\left( {\tau ,\nu } \right)$ before adding noise is the result of the \emph{twisted convolution} between the DD domain transmitted signal ${s_{{\rm{DD}}}}\left( {\tau ,\nu } \right)$ and the DD domain channel response ${h_{{\rm{DD}}}}\left( {\tau ,\nu } \right)$, i.e.,
\begin{align}
{r_{{\rm{DD}}}}\left( \tau, \nu \right) = {h_{{\rm{DD}}}}\left( {\tau ,\nu } \right){ * _\sigma }{s_{{\rm{DD}}}}\left( {\tau ,\nu } \right) \triangleq \int_{ - \infty }^\infty  {\int_{ - \infty }^\infty  {s_{{\rm{DD}}}\left( {\tau  - \tau ',\nu  - \nu '} \right)} } {h_{{\rm{DD}}}}\left( {\tau ',\nu '} \right){e^{j2\pi \nu '\left( {\tau  - \tau '} \right)}}{\rm{d}}\nu '{\rm{d}}\tau ',
\label{DD_twisted_conv_delay_first_Doppler_second}
\end{align}
where ${ * _\sigma }$ is the twisted convolution operator~\cite{MohammedBITSpart1}.

\textbf{Proof}: Let $s_{\rm T}\left( t \right)$ and $r_{\rm T}\left( t \right)$ be the time domain signals corresponding to ${s_{{\rm{DD}}}}\left( {\tau ,\nu } \right)$ and ${r_{{\rm{DD}}}}\left( {\tau ,\nu } \right)$, respectively. Then, according to~\cite{Hadani2017orthogonal}, we have
\begin{align}
{r_{\rm{T}}}\left( t \right) = \int_{ - \infty }^\infty  {\int_{ - \infty }^\infty  {{h_{{\rm{DD}}}}\left( {\tau ,\nu } \right)} } {e^{j2\pi \nu \left( {t - \tau } \right)}}{s_{\rm{T}}}\left( {t - \tau } \right){\rm{d}}\nu {\rm{d}}\tau.
\label{TD_twisted_conv_delay_first_Doppler_second}
\end{align}
Then, by applying the Zak transform to ${r_{\rm{T}}}\left( t \right)$, we obtain
\begin{align}
r_{{\rm{DD}}}\left( {\tau ,\nu } \right) &= \sqrt T \sum\limits_{k =  - \infty }^\infty  {{r_{\rm{T}}}\left( {\tau  + kT} \right){e^{ - j2\pi k\nu T}}} \notag\\
&= \sqrt T \sum\limits_{k =  - \infty }^\infty  {\int_{ - \infty }^\infty  {\int_{ - \infty }^\infty  {{h_{{\rm{DD}}}}\left( {\tau ',\nu '} \right){e^{j2\pi \nu '\left( {\tau  + kT - \tau '} \right)}}{s_{\rm{T}}}\left( {\tau  + kT - \tau '} \right){e^{ - j2\pi k\nu T}}} } {\rm{d}}\nu '{\rm{d}}\tau '} \notag\\
&  = \sqrt T \int_{ - \infty }^\infty  {\int_{ - \infty }^\infty  {\sum\limits_{k =  - \infty }^\infty  {{s_{\rm{T}}}\left( {\tau  - \tau ' + kT} \right)} } } {e^{ - j2\pi k\left( {\nu  - \nu '} \right)T}}{h_{{\rm{DD}}}}\left( {\tau ',\nu '} \right){e^{j2\pi \nu '\left( {\tau  - \tau '} \right)}}{\rm{d}}\nu '{\rm{d}}\tau '\notag\\
& = \int_{ - \infty }^\infty  {\int_{ - \infty }^\infty  {s_{{\rm{DD}}}\left( {\tau  - \tau ',\nu  - \nu '} \right)} } {h_{{\rm{DD}}}}\left( {\tau ',\nu '} \right){e^{j2\pi \nu '\left( {\tau  - \tau '} \right)}}{\rm{d}}\nu '{\rm{d}}\tau '.
\end{align}
\hfill $\blacksquare$

To extract the transmitted information from the received signal $r_{{\rm{DD}}}\left( {\tau ,\nu } \right)$, we apply the matched filtering in the DD domain, yielding
\begin{align}
{Y_{{\rm{DD}}}}\left[ {l,k} \right] = \int_0^T {\int_0^{\frac{1}{T}} {{r_{{\rm{DD}}}}\left( {\tau ,\nu } \right)} } {\left[ {\Phi _{{\rm{DD}}}^{{\tau _l},{\nu _k}}\left( {\tau ,\nu } \right)} \right]^*}{\rm{d}}\nu {\rm{d}}\tau,
\label{DD_received_signal}
\end{align}
where ${{\bf Y}_{{\rm{DD}}}}$ is the set of sufficient statistics used for symbol detection. Based on the above overview of the DD communications, we in the following discuss the design of DD domain basis functions.


\subsection{Delay-Doppler Domain Basis Functions}
Recall that $M$ and $N$ are the numbers of delay and Doppler bins within the fundamental rectangle, respectively.
Thus, a family of equally-spaced DD domain basis functions can be defined by
\begin{align}
{{\bm \Xi} _{{\rm{DD}}}}  \buildrel \Delta \over =  \left\{ {\left. {\Phi _{{\rm{DD}}}^{{\tau _l},{\nu _k}}\left( {\tau ,\nu } \right)} \right|{\tau _l} = l \frac{T}{M},{\nu_k} = k \frac{1}{NT},l \in \left\{ {0,...,M - 1} \right\},k \in \left\{ {0,...,N - 1} \right\}} \right\}. \label{DD_function_set}
\end{align}
Each element in ${\bm \Xi _{{\rm{DD}}}}$ is referred to as a DD domain basis function with delay and Doppler offsets $\tau_l$ and $\nu_k$.
To ensure that the family of basis functions in ${\bm \Xi _{{\rm{DD}}}}$ can be efficiently implemented by a single prototype pulse in practice, we require that
\begin{align}
\Phi _{{\rm{DD}}}^{{\tau _l},{\nu _k}}\left( {\tau ,\nu } \right) = {e^{j2\pi {\nu_k}\left( {\tau  - {\tau _l}} \right)}}\Phi _{{\rm{DD}}}^{0,0}\left( {\tau  - {\tau _l},\nu  - {\nu_k}} \right),\label{DD_basis_offset}
\end{align}
holds for any $\tau\in \left( { - \infty ,\infty } \right)$ and $\nu\in \left( { - \infty ,\infty } \right)$. The phase term ${e^{j2\pi {\nu _k}\left( {\tau  - {\tau _l}} \right)}}$ in~\eqref{DD_basis_offset} comes from the delay-Doppler shifting property in Table~\ref{Zak_properties}, and it is applied to ensure that shifting the DD pulse $\Phi _{{\rm{DD}}}^{0,0}\left( {\tau ,\nu } \right)$ along delay and Doppler axes by ${{\tau _l}}$ and ${{\nu _k}}$ is corresponding to the application of time delay ${{\tau _l}}$ and phase rotations ${e^{j2\pi {\nu_k}\left( {t  - {\tau _l}} \right)}}$ to the time domain equivalent pulse{\footnote{Here, we assume that the signal is first phase-rotated and then time-delayed.}}, i.e., the time domain basis function $\Phi _{{\rm{T}}}^{0,0}\left( t\right)$.
More specifically, applying the inverse Zak transform in~\eqref{Inverse_ZT} to $\Phi _{{\rm{DD}}}^{\tau_l,\nu_k}\left( \tau, \nu \right)$, the time domain basis function $\Phi _{{\rm{T}}}^{\tau_l,\nu_k}\left( t \right)$ can be shown to satisfy
\begin{align}
\Phi _{\rm{T}}^{{\tau _l},{\nu _k}}\left( t \right) = \sqrt T \int_0^{\frac{1}{T}} {\Phi _{{\rm{DD}}}^{{\tau _l},{\nu _k}}\left( {t,\nu } \right){\rm{d}}\nu }  = {e^{j2\pi {\nu_k}\left( {t - {\tau _l}} \right)}}\Phi _{\rm{T}}^{0,0}\left( {t - {\tau _l}} \right).\label{time_basis_offset_property}
\end{align}
Similarly, we can also derive the frequency domain basis function $\Phi _{\rm{F}}^{{\tau _l},{\nu _k}}\left( f \right)$ based on~\eqref{ZT_to_FT}, i.e.,
\begin{align}
\Phi _{\rm{F}}^{{\tau _l},{\nu _k}}\left( f \right) = \frac{1}{{\sqrt T }}\int_0^T {\Phi _{{\rm{DD}}}^{{\tau _l},{\nu _k}}\left( {\tau ,f} \right){e^{ - j2\pi f\tau }}} {\rm{d}}\tau  = {e^{ - j2\pi f{\tau _l}}}\Phi _{\rm{F}}^{0,0}\left( {f - {\nu _k}} \right)
.\label{freq_basis_offset_property}
\end{align}
From~\eqref{time_basis_offset_property} and~\eqref{freq_basis_offset_property}, we observe that the construction in~\eqref{DD_basis_offset} allows the symmetrical treatment for both time and frequency, where the time and frequency domain basis functions are of similar structure, i.e., a carrier that is time/frequency shifted and phase-rotated.
In what follows, we shall refer to~\eqref{DD_basis_offset} as the
\emph{TF-consistency condition} defined in the DD domain, and we say a family of DD domain basis functions are  \emph{TF-consistent} if the functions within ${{\bm \Xi} _{{\rm{DD}}}}$ satisfy~\eqref{DD_basis_offset}. In particular, for TF-consistent family of DD domain basis functions, the following lemma holds.

\textbf{Lemma~3} (\emph{DD shifts for TF-consistent DD basis functions}):
Let $\left( {{\tau _0},{\nu _0}} \right)$ be a pair of delay and Doppler offsets with arbitrary values. Then, for $\Phi _{{\rm{DD}}}^{{\tau _1},{\nu _1}}\left( {\tau ,\nu } \right)$ from ${{\bm \Xi} _{{\rm{DD}}}}$, we have
\begin{align}
{e^{j2\pi {\nu _1}\left( {\tau  - {\tau _1}} \right)}}\Phi _{{\rm{DD}}}^{0,0}\left( {\tau  - {\tau _0} - {\tau _1},\nu  - {\nu _0} - {\nu _1}} \right) = {e^{j2\pi {\nu _1}{\tau _0}}}\Phi _{{\rm{DD}}}^{{\tau _1},{\nu _1}}\left( {\tau  - {\tau _0},\nu  - {\nu _0}} \right) \label{DD_shift_TF_consistent}.
\end{align}

\textbf{Proof}: Define $\tau'=\tau-\tau_0$ and $\nu'=\nu-\nu_0$. Then,
\begin{align}
{e^{j2\pi {\nu _1}\left( {\tau  - {\tau _1}} \right)}}\Phi _{{\rm{DD}}}^{0,0}\left( {\tau  - {\tau _0} - {\tau _1},\nu  - {\nu _0} - {\nu _1}} \right) =& {e^{j2\pi {\nu _1}{\tau _0}}}{e^{j2\pi {\nu _1}\left( {\tau ' - {\tau _1}} \right)}}\Phi _{{\rm{DD}}}^{0,0}\left( {\tau ' - {\tau _1},\nu ' - {\nu _1}} \right)\notag\\
=&{e^{j2\pi {\nu _1}{\tau _0}}}\Phi _{{\rm{DD}}}^{{\tau _1},{\nu _1}}\left( {\tau  - {\tau _0},\nu  - {\nu _0}} \right).
\end{align}\hfill $\blacksquare$

From~\eqref{DD_basis_offset} and~\eqref{DD_shift_TF_consistent}, we notice that the TF-consistency condition states that the basis functions are symmetrically modulated in both time and frequency following the same operations. More importantly, the following theorem shows the important connection between TF consistency and ambiguity function.

\textbf{Theorem~2} (\emph{TF-consistency vs. Ambiguity Function}):
Let $\left(\tau_1, \nu_1\right)$ and $\left(\tau_2, \nu_2\right)$ be two pairs of arbitrary delay and Doppler offsets. Then, the following holds
\begin{align}
&\int_0^T {\int_0^{\frac{1}{T}} {{e^{j2\pi {\nu _2}\left( {\tau  - {\tau _2}} \right)}}{{\cal Z}_x}\left( {\tau  - {\tau _2},\nu  - {\nu _2}} \right){e^{ - j2\pi {\nu _1}\left( {\tau  - {\tau _1}} \right)}}{\cal Z}_x^*\left( {\tau  - {\tau _1},\nu  - {\nu _1}} \right){\rm{d}}\nu {\rm{d}}\tau } }  \notag\\
=& {e^{j2\pi {\nu _2}\left( {{\tau _1} - {\tau _2}} \right)}}{A_x}\left( {{\tau _1} - {\tau _2},{\nu _1} - {\nu _2}} \right),
\label{DD_basis_function_00_vs_AF}
\end{align}
where ${A_{x}}\left( {{\Delta \tau},{\Delta \nu}} \right)$ denotes the auto-ambiguity function of $x\left( {t } \right)$ with respect to the delay offset ${\Delta \tau}$ and Doppler offset ${\Delta \nu}$.
Particularly, for ${\Phi _{{\rm{DD}}}^{{\tau _1},{\nu _1}}\left( {\tau ,\nu } \right)}$ and ${\Phi _{{\rm{DD}}}^{{\tau _2},{\nu _2}}\left( {\tau ,\nu } \right)}$ belonging to ${{\bm \Xi} _{{\rm{DD}}}}$,~\eqref{DD_basis_function_00_vs_AF} suggests
\begin{align}
\int_0^T {\int_0^{\frac{1}{T}} {\Phi _{{\rm{DD}}}^{{\tau _2},{\nu _2}}\left( {\tau ,\nu } \right){{\left( {\Phi _{{\rm{DD}}}^{{\tau _1},{\nu _1}}\left( {\tau ,\nu } \right)} \right)}^*}{\rm{d}}\nu {\rm{d}}\tau } }
={e^{j2\pi {\nu _2}\left( {{\tau _1} - {\tau _2}} \right)}}{A_\Phi }\left( {{\tau _1} - {\tau _2},{\nu _1} - {\nu _2}} \right),
\label{DD_MF_basis_functions}
\end{align}
where ${A_{\Phi}}\left( {{\Delta \tau},{\Delta \nu}} \right)$ denotes the auto-ambiguity function of $\Phi _{\rm{T}}^{0,0}\left( {t } \right)$.

\textbf{Proof}: The theorem can be straightforwardly derived based on Lemma~2 and the delay-Doppler shifting property in Table~\ref{Zak_properties}. \hfill $\blacksquare$

Notice that~\eqref{DD_MF_basis_functions} is of the form of DD domain matched-filtering.
Essentially, the property in~\eqref{DD_MF_basis_functions} suggests that the DD domain signal transmission with a family of TF-consistent DD domain basis functions can be characterized the ambiguity function of $\Phi _{\rm{T}}^{0,0}\left( {t } \right)$. Given above, we shall view ${\Phi _{{\rm{DD}}}^{{0},{0}}\left( {\tau ,\nu } \right)}$ as the DD domain \emph{prototype pulse} for DD pulse shaping.

\textbf{Remark~1}: We highlight that the above description aligns with the DD domain modulation discussed in~\cite{MohammedBITSpart1,Mohammed2023part2},
where the authors have shown that the twisted-convolution discussed in Theorem~1 is the DD domain basic operations characterizing the DD pulse shaping, signal transmission, and the matched-filtering.
In comparison to this descriptions, the description above highlights the intrinsic
connections among different DD domain basis functions from the view point of communication theory by explicitly offering the mathematical description of carrier pulses with respect to each information symbol. In the following subsections, we will reveal the important insights of such constructions for practical communication implementations.

\subsection{Constructing Delay-Doppler Domain Basis Functions}
\begin{figure}
\centering
\includegraphics[width=0.8\textwidth]{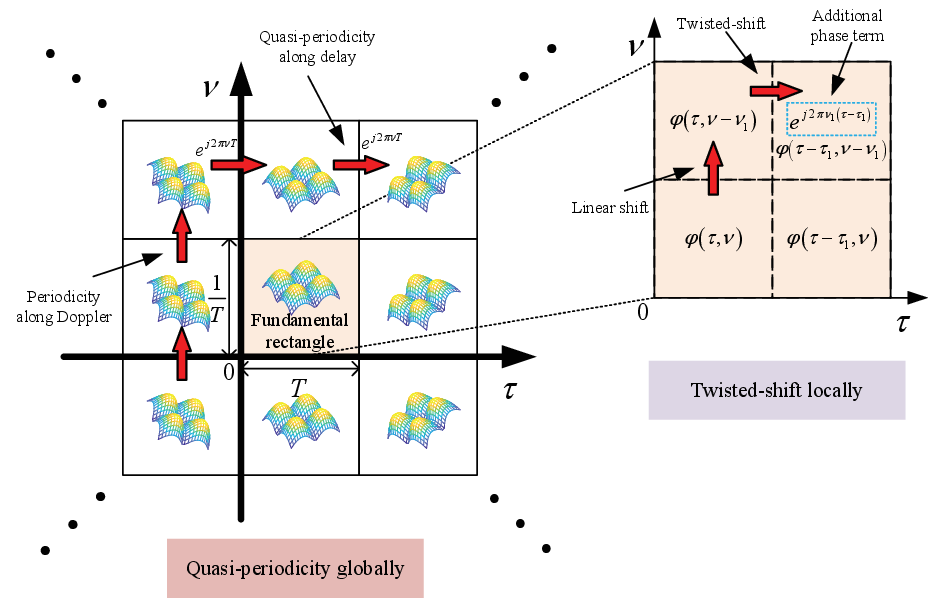}
\caption{DD domain basis functions exhibit quasi-periodicity globally, which is constructed by locally twisted-shifting the atom pulse $\varphi \left( {\tau ,\nu } \right)$. Here, we assume $M=N=2$.}
\label{DD_basis_fig}
\centering
\end{figure}

Note that any DD domain signal satisfying the quasi-periodicity property can be sufficiently characterized by the corresponding response in the fundamental rectangle. Thus, we are motivated to construct the DD domain basis function by extending the ``atom pulse'' in the fundamental rectangle following the  quasi-periodicity.
Let $\varphi \left( {\tau ,\nu } \right)$ be the atom pulse, whose support is the fundamental rectangle. According to Lemma~1, ${\Phi _{{\rm{DD}}}^{0,0}\left( {\tau ,\nu } \right)}$ can then be obtained by quasi-periodically extending $\varphi \left( {\tau ,\nu } \right)$~\cite{MohammedBITSpart1,Mohammed2023part2}, such as
\begin{align}
\Phi _{{\rm{DD}}}^{0,0}\left( {\tau ,\nu } \right) \buildrel \Delta \over = \sum\limits_{n =  - \infty }^\infty  {\sum\limits_{m =  - \infty }^\infty  {\varphi \left( {\tau  - nT,\nu  - \frac{m}{T}} \right)} } {e^{j2\pi n\nu T}}.\label{DD_basis_function_construction}
\end{align}
By substituting~\eqref{DD_basis_function_construction} into~\eqref{DD_basis_offset}, we obtain
\begin{align}
\Phi _{{\rm{DD}}}^{{\tau _l},{\nu _k}}\left( {\tau ,\nu } \right) &= \sum\limits_{n =  - \infty }^\infty  {\sum\limits_{m =  - \infty }^\infty  {{e^{j2\pi {\nu _k}\left( {\tau  - {\tau _l}} \right)}}\varphi \left( {\tau  - {\tau _l} - nT,\nu  - {\nu _k} - \frac{m}{T}} \right)} } {e^{j2\pi n\left( {\nu  - {\nu _k}} \right)T}}.\label{DD_basis_function_offset_construction}
\end{align}
We summarize the properties of the DD domain basis functions in Fig.~\ref{DD_basis_fig}, where we assume that $M=N=2$ such that there are $MN=4$ DD domain basis functions in the fundamental rectangle. We observe that the DD domain basis functions exhibit quasi-periodicity globally, and its constructed by locally twisted-shifting the atom pulse $\varphi \left( {\tau ,\nu } \right)$.
Specifically, the quasi-periodicity aligns with the property of the Zak transform, which characterizes the global structure of the DD domain basis functions across infinite numbers of regions with the size of the fundamental rectangle. On the other hand, the twisted-shift aligns with the TF-consistency condition, characterizing the local structure of the DD domain basis functions across $M$ delay bins and $N$ Doppler bins.

\begin{figure}
\centering
\includegraphics[width=0.6\textwidth]{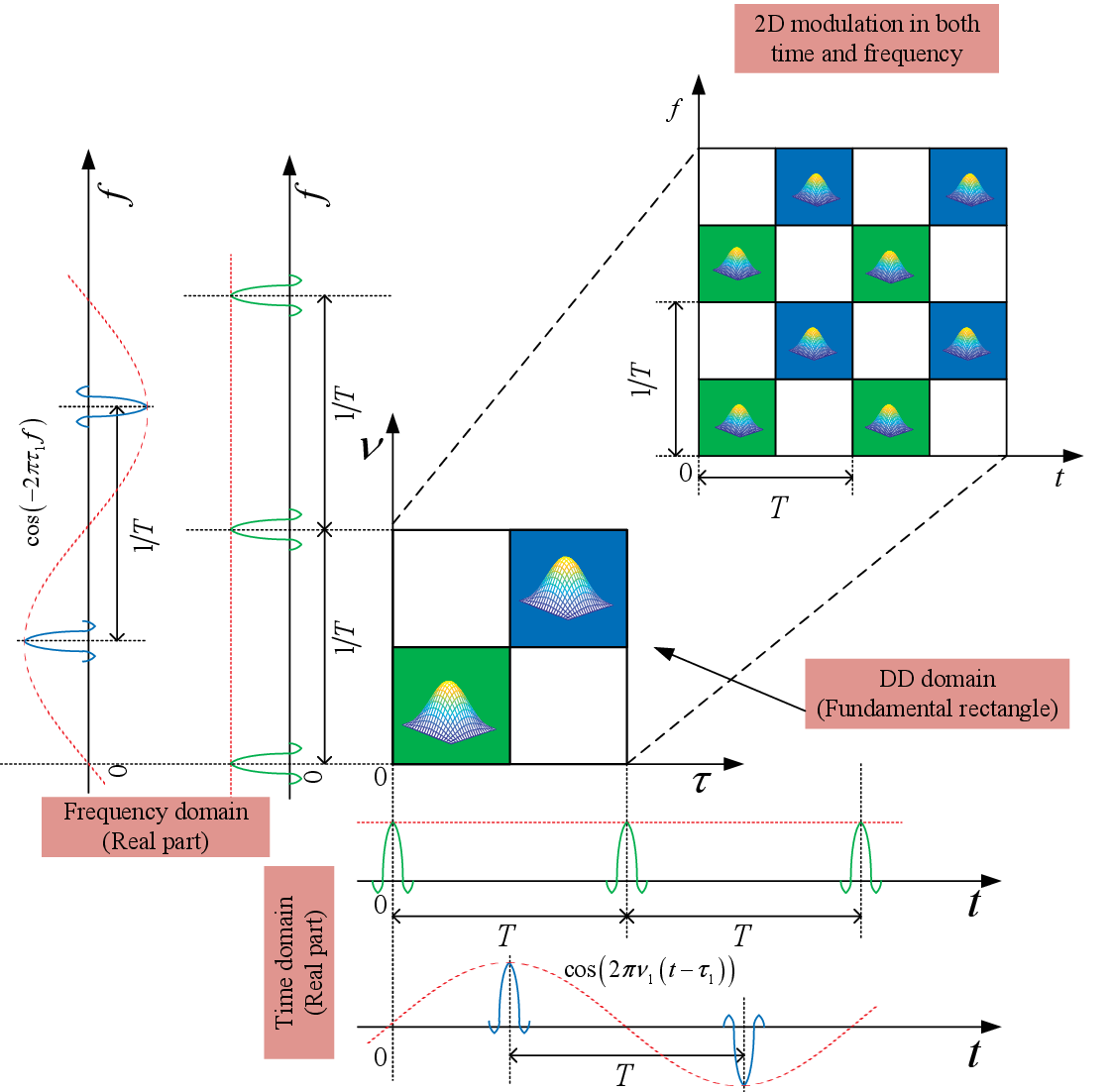}
\caption{DD domain basis functions and equivalent representations in time and frequency domains. Here, the time and frequency pulses are with the same colors as the corresponding DD domain grid. In the figure, we assume $M=N=2$.}
\label{DD_basis_T_F_fig}
\centering
\end{figure}

The significance of the characteristics of the DD domain basis functions have been discussed~\cite{MohammedBITSpart1,Mohammed2023part2}.
But here we propose a different viewpoint from the DD domain pulse structures.
We argue that the global and local characteristics allow the DD domain basis functions have direct time domain and frequency domain interpretations via exploiting the properties of the Zak transform, e.g.,~\eqref{Inverse_ZT} and~\eqref{ZT_to_FT}. Specifically, by substituting~\eqref{DD_basis_function_construction} into~\eqref{time_basis_offset_property}, we obtain the time domain basis function by
\begin{align}
\Phi _{\rm{T}}^{0,0}\left( t \right) = \sqrt T \int_0^{\frac{1}{T}} {\sum\limits_{n =  - \infty }^\infty  {\sum\limits_{m =  - \infty }^\infty  {\varphi \left( {t - nT,\nu  - \frac{m}{T}} \right)} } {e^{j2\pi n\nu T}}{\rm{d}}\nu }  = \sqrt T \sum\limits_{n =  - \infty }^\infty  {\int_{ - \infty }^\infty  {\varphi \left( {t - nT,\nu } \right){e^{j2\pi n\nu T}}{\rm{d}}\nu } }   . \label{time_basis_function_00}
\end{align}
Similarly, by substituting~\eqref{DD_basis_function_offset_construction} into~\eqref{ZT_to_FT}, we obtain the frequency domain basis function by
\begin{align}
\Phi _{\rm F}^{0,0}\left( f \right) =& \frac{1}{{\sqrt T }}\int_0^T {\sum\limits_{n =  - \infty }^\infty  {\sum\limits_{m =  - \infty }^\infty  {\varphi \left( {\tau  - nT,f - \frac{m}{T}} \right)} } {e^{j2\pi nfT}}{e^{ - j2\pi f\tau }}{\rm{d}}\tau } \notag\\
=& \frac{1}{{\sqrt T }}\sum\limits_{m =  - \infty }^\infty  {\int_{ - \infty }^\infty  {\varphi \left( {\tau ,f - \frac{m}{T}} \right){e^{ - j2\pi f\tau }}{\rm{d}}\tau } } .
\label{frequency_basis_function_00}
\end{align}
Furthermore, according to~\eqref{time_basis_offset_property} and~\eqref{ZT_to_FT}, we have
\begin{align}
\Phi _{\rm{T}}^{{\tau _l},{\nu _k}}\left( t \right) ={e^{j2\pi {\nu_k}\left( {t - {\tau _l}} \right)}}\Phi _{\rm{T}}^{0,0}\left( {t - {\tau _l}} \right)= \sqrt T {e^{j2\pi {\nu_k}\left( {t - {\tau _l}} \right)}}\sum\limits_{n =  - \infty }^\infty  {\int_{-\infty}^\infty {\varphi \left( {t - {\tau _l}- nT,\nu } \right){e^{j2\pi n\nu T}}{\rm{d}}\nu } } ,
\label{time_basis_function_lk}
\end{align}
and
\begin{align}
\Phi _{\rm{F}}^{{\tau _l},{\nu _k}}\left( f \right) = {e^{ - j2\pi f{\tau _l}}}\Phi _{\rm{F}}^{0,0}\left( {f - {\nu _k}} \right) = \frac{1}{{\sqrt T }}{e^{ - j2\pi f{\tau _l}}}\sum\limits_{m =  - \infty }^\infty  {\int_{-\infty}^\infty {\varphi \left( {\tau ,f - {\nu _k}- \frac{m}{T}} \right){e^{ - j2\pi f\tau }}{\rm{d}}\tau } } .
\label{frequency_basis_function_lk}
\end{align}
Based on~\eqref{time_basis_function_lk} and~\eqref{frequency_basis_function_lk}, we observe that
the DD domain basis functions can be understood as a mixture of the time domain and frequency domain pulses/tones~\cite{MohammedBITSpart1,Mohammed2023part2}, by noticing that integrals in~\eqref{time_basis_function_lk} and~\eqref{frequency_basis_function_lk} result in purely one-dimensional (1D) time and frequency signals. As an example, we demonstrate the basis functions in different domains in Fig.~\ref{DD_basis_T_F_fig}, where we assume $M=N=2$ and mark the corresponding time and frequency pulses the same colors as those in the DD grids. As shown in the figure, the DD domain basis function becomes 1D  pulsone in either time or frequency, while showing a particular response pattern according to the delay and Doppler offsets. Particularly, we have the following observations:
\begin{itemize}
  \item \textbf{DD domain global properties characterize the time and frequency periodicity}: The DD domain global characteristics are translated into the summation and integral terms in~\eqref{time_basis_function_lk} and~\eqref{frequency_basis_function_lk}, which leads to a train of pulses in the time and frequency domains that are apart in time by $T$ and apart in frequency by $\frac{1}{T}$, respectively.
  \item \textbf{DD domain local properties characterize time and frequency tones}: The DD domain local characteristics are translated into the phase terms (signal tones) ${e^{j2\pi {\nu_k}\left( {t - {\tau _l}} \right)}}$ and ${e^{ - j2\pi f{\tau _l}}}$ in~\eqref{time_basis_function_lk} and~\eqref{frequency_basis_function_lk}.
  \item \textbf{Time and frequency spreading}: The DD domain basis function is spread in both time and frequency simultaneously following a periodic manner with respect to $T$ and $\frac{1}{T}$, leading to a potential of achieving full channel diversity{\footnote{We highlight here that the combination of time domain and frequency domain is not the commonly known time-frequency domain, where conventional OFDM multiplexes the information symbol.}}.
      Specifically, the special signal structure of~\eqref{time_basis_function_lk} and~\eqref{frequency_basis_function_lk} is known as the \emph{pulsone}~\cite{MohammedBITSpart1,Mohammed2023part2}, which is essentially a \emph{pulse train} modulated by a \emph{complex tone}.
  \item \textbf{Time and frequency limiting cases}: By letting $T \to \infty$, the time domain basis functions are separated only by the time offset $\tau_l$, yielding a pure time division multiplexing (TDM)-type of signaling; By letting $T \to 0$, the frequency domain basis functions are separated only by the frequency offset $\nu_k$, yielding a pure frequency division multiplexing (FDM)-type of signaling.
  \end{itemize}

\section{Properties of DD Domain Basis Function and It's Truncation}
In this section, we study the properties of DD domain basis functions based on the fundamental understanding from the previous sections.
Perhaps, the \emph{ideal} DD domain basis functions may be a set of quasi-periodically extended delta pulses (distributions), i.e., $\varphi \left( {\tau ,\nu } \right) = \delta \left( \tau  \right)\delta \left( \nu  \right)$, which are ideally localized in the fundamental rectangle and thereby minimizes the interference among information symbols. It should be noted that the well-known \emph{Heisenberg's uncertainty principle} forbids the existence of fully localized pulses in the TF domain~\cite{mallat1999wavelet}. However, the Heisenberg's uncertainty principle does not apply to the Zak domain directly. More specifically, we shall highlight that the ideal DD domain basis functions are only fully localized in the fundamental rectangle, while are in fact quasi-periodic in the whole DD domain globally according to~\eqref{DD_basis_function_offset_construction}.
Particularly, by considering
\begin{align}
\Phi _{{\rm{DD}}}^{{\tau _l},{\nu _k}}\left( {\tau ,\nu } \right) = \sum\limits_{n =  - \infty }^\infty  {\sum\limits_{m =  - \infty }^\infty  {{e^{j2\pi {\nu _k}\left( {\tau  - {\tau _l}} \right)}}\delta \left( {\tau  - {\tau _l} - nT} \right)\delta \left( {\nu  - {\nu _k} - \frac{m}{T}} \right)} } {e^{j2\pi n\left( {\nu  - {\nu _k}} \right)T}},
\label{DD_basis_localized}
\end{align}
we obtain
\begin{align}
\Phi _{\rm{T}}^{{\tau _l},{\nu _k}}\left( t \right) = \sqrt T {e^{j2\pi {\nu_k}\left( {t - {\tau _l}} \right)}}\sum\limits_{n =  - \infty }^\infty  {\delta \left( {t - {\tau _l} - nT} \right)} ,
\label{time_basis_localized}
\end{align}
and
\begin{align}
\Phi _{\rm{F}}^{{\tau _l},{\nu _k}}\left( f \right) = \frac{1}{{\sqrt T }}{e^{ - j2\pi f{\tau _l}}}\sum\limits_{m =  - \infty }^\infty  {\delta \left( {f - {\nu _k} - \frac{m}{T}} \right)}
\label{freq_basis_localized}
\end{align}
respectively. Furthermore, we have
\begin{align}
{A_\Phi }\left( {{\tau _1} - {\tau _2},{\nu _1} - {\nu _2}} \right)
= \sum\limits_{n =  - \infty }^\infty  {\sum\limits_{m =  - \infty }^\infty  \delta  } \left( {{\tau _1} - {\tau _2} - nT} \right)\delta \left( {{\nu _1} - {\nu _2} - \frac{m}{T}} \right),
  \label{AF_fully_localized}
\end{align}
whose detailed derivation is given in Appendix~A. In DD communications, the ambiguity function of the form of~\eqref{AF_fully_localized} minimizes the potential interference among different information symbols, but it only exists in theory but not in practice, because~\eqref{time_basis_localized} and~\eqref{freq_basis_localized} clearly suggest infinite time and frequency resources.
Therefore, we propose to apply practical filters and windows to limit the occupied TF resources of~\eqref{time_basis_localized} and~\eqref{freq_basis_localized}.

\subsection{Time-Frequency Consistent Filtering and Windowing}
The idea of applying filtering and windowing for limiting the TF resources is straightforward, and previous implementation on OTFS based on this appears in~\cite{mohammed2021derivation}. However, what is not obvious and easy to be overlooked is the TF-consistency condition. We have shown in~\eqref{DD_basis_function_00_vs_AF} that the TF-consistency condition directly connects the DD domain matched-filtering and the ambiguity function. Consequently, filtering or windowing that does not align with the TF-consistency will break the DD domain integrity, and therefore degrades the communication performance.

In the following, we study the time domain TF-consistent filtering/windowing for the sake of practical implementation.
For a family of DD domain basis functions, we shall define the TF-consistent filtering/windowing in the following Proposition.

\textbf{Proposition~1} (\emph{TF-Consistent Filtering/Windowing}):
Define a family of DD domain basis functions ${{\bf{\Xi }}_{{\rm{DD}}}}$ that are delay and Doppler shifted with respect to a prototype pulse $\Phi _{{\rm{DD}}}^{0,0}\left( {\tau ,\nu } \right)$ in a TF-consistent manner, i.e.,~\eqref{DD_basis_offset}.
Define another family of DD domain basis functions ${{\bf{\tilde \Xi }}_{{\rm{DD}}}}$ that are obtained by time domain filtering or windowing each corresponding time domain basis function from
${{\bf{\Xi }}_{{\rm{DD}}}}$. We call the filtering/windowing is TF-consistent if and only if  ${{ \bf{\tilde \Xi }}_{{\rm{DD}}}}$ is TF-consistent.

To study the operational meaning of the TF-consistent filtering/windowing. Let us define an arbitrary time domain function $x\left(t\right)$ as the filter/window function.
Let us first study the TF-consistent time domain filtering. We define $\tilde \Phi _{{\rm{DD}}}^{0,0}\left( {\tau ,\nu } \right) = \frac{1}{{\sqrt T }}\int_0^T {\Phi _{{\rm{DD}}}^{0,0}\left( {\tau  - \tau ',\nu } \right){{\cal Z}_x}\left( {\tau ',\nu } \right)} {\rm{d}}\tau '$, i.e., $\tilde \Phi _{\rm{T}}^{0,0}\left( t \right) = \Phi _{\rm{T}}^{0,0}\left( t \right) * x\left( t \right)$.
Immediately from Proposition~1, we shall write
\begin{align}
{e^{j2\pi {\nu _0}\left( {\tau  - {\tau _0}} \right)}}\tilde \Phi _{{\rm{DD}}}^{0,0}\left( {\tau  - {\tau _0},\nu  - {\nu _0}} \right) = & \frac{1}{{\sqrt T }}\int_0^T {{e^{j2\pi {\nu _0}\left( {\tau  - {\tau _0}} \right)}}\Phi _{{\rm{DD}}}^{0,0}\left( {\tau  - \tau ' - {\tau _0},\nu  - {\nu _0}} \right){{\cal Z}_x}\left( {\tau ',\nu  - {\nu _0}} \right)} {\rm{d}}\tau '\notag\\
=&\frac{1}{{\sqrt T }}\int_0^T {{e^{j2\pi {\nu _0}\tau '}}\Phi _{{\rm{DD}}}^{{\tau _0},{\nu _0}}\left( {\tau  - \tau ',\nu } \right){{\cal Z}_x}\left( {\tau ',\nu  - {\nu _0}} \right)} {\rm{d}}\tau '.
\label{TF_consistent_filtering_DD}
\end{align}
where~\eqref{TF_consistent_filtering_DD} comes from~\eqref{DD_shift_TF_consistent}.
By taking the inverse Zak transform of~\eqref{TF_consistent_filtering_DD}, we obtain
\begin{align}
\tilde \Phi _{\rm{T}}^{{\tau _0},{\nu _0}}\left( t \right) = \Phi _{\rm{T}}^{{\tau _0},{\nu _0}}\left( t \right)* \left( {{e^{j2\pi {\nu _0}t}}x\left( t \right)} \right).
\label{TF_consistent_filtering_time}
\end{align}
For the time domain windowing, let us define $\tilde \Phi _{{\rm{DD}}}^{0,0}\left( {\tau ,\nu } \right) = \sqrt T \int_0^{\frac{1}{T}} {\Phi _{{\rm{DD}}}^{0,0}\left( {\tau ,\nu  - \nu '} \right){{\cal Z}_x}\left( {\tau ,\nu '} \right)} {\rm{d}}\nu '$, i.e., $\tilde \Phi _{\rm{T}}^{0,0}\left( t \right) = \Phi _{\rm{T}}^{0,0}\left( t \right)x\left( t \right)$.
Again from Proposition~1, we shall write
\begin{align}
{e^{j2\pi {\nu _0}\left( {\tau  - {\tau _0}} \right)}}\tilde \Phi _{{\rm{DD}}}^{0,0}\left( {\tau  - {\tau _0},\nu  - {\nu _0}} \right) =& \sqrt T \int_0^{\frac{1}{T}} {{e^{j2\pi {\nu _0}\left( {\tau  - {\tau _0}} \right)}}\Phi _{{\rm{DD}}}^{0,0}\left( {\tau  - {\tau _0},\nu  - {\nu _0} - \nu '} \right){{\cal Z}_x}\left( {\tau  - {\tau _0},\nu '} \right)} {\rm{d}}\nu '\notag\\
=& \sqrt T \int_0^{\frac{1}{T}} {\Phi _{{\rm{DD}}}^{{\tau _0},{\nu _0}}\left( {\tau ,\nu  - \nu '} \right){{\cal Z}_x}\left( {\tau  - {\tau _0},\nu '} \right)} {\rm{d}}\nu ',
\label{TF_consistent_windowing_DD}
\end{align}
where~\eqref{TF_consistent_windowing_DD} comes from~\eqref{DD_shift_TF_consistent}.
By taking the inverse Zak transform of~\eqref{TF_consistent_windowing_DD}, we obtain
\begin{align}
\tilde \Phi _{\rm{T}}^{{\tau _0},{\nu _0}}\left( t \right) = \Phi _{\rm{T}}^{{\tau _0},{\nu _0}}\left( t \right)x\left( {t - {\tau _0}} \right).
\label{TF_consistent_windowing_time}
\end{align}
It can be seen from that~\eqref{TF_consistent_filtering_time} and~\eqref{TF_consistent_windowing_time} suggest that TF-consistency condition can be preserved if the filter/window function is frequency/time shifted according to the Doppler/delay offsets. In fact, this result is not unexpected. Note that both ${{\bf{\Xi }}_{{\rm{DD}}}}$ consists of a family of pulses with different Doppler/delay offsets. Therefore, to make sure each pulse undergoes the exactly same effect of filtering/windowing, it is necessary to also adapt the filter/window response according to the Doppler/delay offsets.
\subsection{DD Domain TF-Consistent Pulse Shaping with Truncated Periodic Signals}
We now discuss the DD domain basis function after TF-consistent filtering and windowing. Considering the practical implementation of DD communications, we restrict ourselves to only consider the case that the DD domain basis function is firstly truncated in frequency and then truncated in time following the TF-consistency condition. As a result, we shall notice that such a truncation produces a roughly time and frequency limited signal. Let $\tilde \Phi _{{\rm{DD}}}^{{\tau _0},{\nu _0}}\left( {\tau ,\nu } \right)$ be the truncated DD domain basis function, whose equivalent time domain representation satisfies
\begin{align}
\tilde \Phi _{\rm{T}}^{{\tau _0},{\nu _0}}\left( t \right) \buildrel \Delta \over = \left\{ {\Phi _{\rm{T}}^{{\tau _0},{\nu _0}}\left( t \right) * \left( {{e^{j2\pi {\nu _0}t}}{\mathsf {FW}_{\rm T}}\left( t \right)} \right)} \right\}{\mathsf {TW}_{\rm T}}\left( {t - {\tau _0}} \right),
\label{DD_basis_truncated_T}
\end{align}
where ${\mathsf {FW}}_{\rm T}\left( t \right)$ and ${\mathsf {TW}}_{\rm T}\left( t \right)$ are the time domain representations of arbitrary frequency domain and time domain windows, respectively.
Corresponding to~\eqref{DD_basis_truncated_T}, the frequency domain basis function after truncation is given by
\begin{align}
\tilde \Phi _{\rm{F}}^{{\tau _0},{\nu _0}}\left( f \right) \buildrel \Delta \over = \left\{ {\Phi _{\rm{F}}^{{\tau _0},{\nu _0}}\left( f \right){\mathsf {FW}}_{\rm F}\left( {f - {\nu _0}} \right)} \right\} * \left( {{e^{j2\pi f{\tau _0}}}{\mathsf {TW}}_{\rm F}\left( f \right)} \right).
\label{DD_basis_truncated_F}
\end{align}
Furthermore, according to the properties in Table~\ref{Zak_properties}, we obtain the truncated DD domain basis function as
\begin{align}
\tilde \Phi _{{\rm{DD}}}^{{\tau _l},{\nu _k}}\left( {\tau ,\nu } \right) = \int_0^{\frac{1}{T}} {\int_0^T {\Phi _{{\rm{DD}}}^{{\tau _l},{\nu _k}}\left( {\tau  - \tau ',\nu  - \nu '} \right){\mathsf {FW}}_{{\rm{DD}}}}\left( {\tau ',\nu  - \nu ' - {\nu _k}} \right){e^{j2\pi {\nu _k}\tau '}}}  {\mathsf {TW}}_{{\rm{DD}}}\left( {\tau  - {\tau _l},\nu '} \right){\rm{d}}\tau '{\rm{d}}\nu '.
\label{DD_basis_truncated_DD}
\end{align}
By substituting~\eqref{DD_basis_localized} into~\eqref{DD_basis_truncated_DD} and considering ${\tau _l}={\nu _k}=0$, we obtain
\begin{align}
&\tilde \Phi _{{\rm{DD}}}^{0,0}\left( {\tau ,\nu } \right)\notag\\
= & \int_0^{\frac{1}{T}} {\int_0^T {\sum\limits_{n =  - \infty }^\infty  {\sum\limits_{m =  - \infty }^\infty  {\delta \left( {\tau  - \tau ' - nT} \right)} } \delta \left( {\nu  - \nu ' - \frac{m}{T}} \right){e^{j2\pi n\left( {\nu  - \nu '} \right)T}}{\mathsf {FW}}_{{\rm{DD}}}\left( {\tau ',\nu  - \nu '} \right)} } {\mathsf {TW}}_{{\rm{DD}}}\left( {\tau ,\nu '} \right){\rm{d}}\tau '{\rm{d}}\nu '\notag\\
=&\sum\limits_{n =  - \infty }^\infty \! {\sum\limits_{m =  - \infty }^\infty \! {\int_{\frac{m}{T}}^{\frac{{m + 1}}{T}}\! {\int_{nT}^{\left( {n + 1} \right)T} {\!\!\delta \left( {\tau \! - \!\tau '} \right)\delta \left( {\nu  \!- \!\nu '} \right){e^{j2\pi n\left( {\nu  - \nu '} \right)T}}{\mathsf {FW}}_{{\rm{DD}}}\left( {\tau ' - nT,\nu  - \nu ' - \frac{m}{T}} \right)} } } } {\mathsf {TW}}_{{\rm{DD}}}\left( {\tau ,\nu ' - \frac{m}{T}} \right){\rm{d}}\tau '{\rm{d}}\nu '\notag\\
=&\int_{ - \infty }^\infty  {\int_{ - \infty }^\infty  {\delta \left( {\tau  - \tau '} \right)\delta \left( {\nu  - \nu '} \right){\mathsf {FW}}_{{\rm{DD}}}\left( {\tau ',\nu  - \nu '} \right)} } {\mathsf {TW}}_{{\rm{DD}}}\left( {\tau ,\nu '} \right){\rm{d}}\tau '{\rm{d}}\nu '\notag\\
=& {\mathsf {FW}}_{{\rm{DD}}}\left( {\tau ,0} \right){\mathsf {TW}}_{{\rm{DD}}}\left( {\tau ,\nu } \right).
\label{DD_basis_truncated_DD}
\end{align}
It is not surprising to see from~\eqref{DD_basis_truncated_DD} that the truncated DD domain basis function is fully determined by the DD domain representations of the time and frequency windows. More specifically, we notice that the the resultant truncated DD domain basis function coincide with the shape of the time domain window along the Doppler axis, while its response along the delay axis is determined jointly by both the time and frequency domain windows.

To further discuss the insight based on~\eqref{DD_basis_truncated_DD}, let us consider time and frequency windows with specific constraints.
We consider the time domain window has a finite time duration from $t \in \left[ {0,\tilde NT} \right]$, while the frequency domain window has a finite bandwidth $f \in \left[ {0,\frac{{\tilde M}}{T}} \right]$, respectively, where $\tilde N \ge N$ and $\tilde M \ge M$. Notice that the Zak transform involves periodic summations in time or frequency as shown in~\eqref{Zak_Transform_def} and~\eqref{ZT_def_FT}. We are motivated to consider \emph{periodic windows} in time and frequency for DD basis function truncation.
Specifically, we call a time domain window ${\mathsf {TW}}_{\rm{T}}\left( t \right)$ $T$-periodic if ${\mathsf {TW}}_{\rm{T}}\left( t \right)={\mathsf {TW}}_{\rm{T}}\left( t+T \right)$, for $t \in \left[ {0,\left(\tilde N-1\right)T} \right]$. Similarly, a frequency domain $\frac{{\tilde M}}{T}$-periodic window satisfies $f \in \left[ {0,\frac{{\tilde M-1}}{T}} \right]$.
With time and frequency periodic windows described above,~\eqref{DD_basis_truncated_DD} is further simplified by
\begin{align}
\tilde \Phi _{{\rm{DD}}}^{0,0}\left( {\tau ,\nu } \right) =& \frac{1}{{\sqrt T }}\sum\limits_{l = 0}^{\tilde M - 1} {{\mathsf {FW}}_{\rm{F}}\left( {\frac{l}{T}} \right)} {e^{j2\pi l\frac{\tau }{T}}}\sqrt T \sum\limits_{k = 0}^{\tilde N - 1} {{\mathsf {TW}}_{\rm{T}}} \left( {\tau  + kT} \right){e^{ - j2\pi k\nu T}}\notag\\
=&{\mathsf {FW}}_{\rm{F}}\left( 0 \right){\mathsf {TW}}_{\rm{T}}\left( \tau  \right)\sum\limits_{l = 0}^{\tilde M - 1} {{e^{j2\pi l\frac{\tau }{T}}}} \sum\limits_{k = 0}^{\tilde N - 1} {{e^{ - j2\pi k\nu T}}} \label{DD_basis_truncated_DD_periodic_window_der1}\\
=&{\mathsf {FW}}_{\rm{F}}\left( 0 \right){\mathsf {TW}}_{\rm{T}}\left( \tau  \right){e^{j\pi \left( {\tilde M - 1} \right)\frac{\tau }{T}}}{e^{-j\pi \left( {\tilde N - 1} \right)\nu T}}\frac{{\sin \left( {\pi \tilde M\frac{\tau }{T}} \right)}}{{\sin \left( {\pi \frac{\tau }{T}} \right)}}\frac{{\sin \left( {\pi \tilde N\nu T} \right)}}{{\sin \left( {\pi \nu T} \right)}},
\label{DD_basis_truncated_DD_periodic_window}
\end{align}
where the summations in~\eqref{DD_basis_truncated_DD_periodic_window_der1} are commonly referred to as the Dirichlet kernel.
Furthermore, it is interesting to notice from~\eqref{DD_basis_truncated_DD_periodic_window} that the signal strength of  $\tilde \Phi _{{\rm{DD}}}^{0,0}\left( {\tau ,\nu } \right)$ is dominated by the terms $\frac{{\sin \left( {\pi \tilde M\frac{\tau }{T}} \right)}}{{\sin \left( {\pi \frac{\tau }{T}} \right)}}$ and $\frac{{\sin \left( {\pi \tilde N\nu T} \right)}}{{\sin \left( {\pi \nu T} \right)}}$.
In fact, signals of the form of $\frac{{\sin \left( {\pi \tilde M\frac{\tau }{T}} \right)}}{{\sin \left( {\pi \frac{\tau }{T}} \right)}}$ and $\frac{{\sin \left( {\pi \tilde N\nu T} \right)}}{{\sin \left( {\pi \nu T} \right)}}$ are commonly referred to as \emph{aliased sinc functions}, or for short, asinc functions, in the literature, which are a special type of quasi-orthogonal signals with respect to the intervals $\frac{T}{\tilde M}$ and $\frac{1}{\tilde NT}$, respectively.
For $\frac{{\sin \left( {\pi  N\nu T} \right)}}{{\sin \left( {\pi \nu T} \right)}}$, its value is zero if $\nu$ is integer multiple of the quasi-orthogonality period $\frac{1}{NT}$, except for the case where $\nu=\frac{lN}{NT}$ with any integer $l$.
Therefore, the DD domain basis function of the form~\eqref{DD_basis_truncated_DD_periodic_window} is desirable in the sense that it naturally provides sufficient orthogonality in the fundamental rectangle for $\tau  \in \left[ {0,T} \right)$ and $\nu  \in \left[ {0,\frac{1}{T}} \right)$. Moreover,~\eqref{DD_basis_truncated_DD_periodic_window} suggests that any periodic windows essentially lead to very similar $\tilde \Phi _{{\rm{DD}}}^{0,0}\left( {\tau ,\nu } \right)$. This observation indicates that DD Nyquist communications can be achieved by simply applying periodic windows without the need of sophisticated pulse design.

For completeness, let us also discuss the time and frequency domain basis functions after such windowing. Note that the time and frequency domain windows are of finite duration. Therefore, their signal responses are invariant after multiplying a rectangular window function in the corresponding domain, such as
\begin{align}
\tilde \Phi _{\rm{T}}^{{\tau _0},{\nu _0}}\left( t \right) =& \frac{{\tilde M}}{T}\left\{ {\Phi _{\rm{T}}^{{\tau _0},{\nu _0}}\left( t \right) * \left( {{e^{j2\pi {\nu _0}t}}{e^{j\pi \frac{{\tilde M}}{T}t}}{\mathsf {FW}}_{\rm{T}}\left( t \right){\rm{sinc}}\left( {\frac{{\tilde M}}{T}t} \right)} \right)} \right\}{\mathsf {TW}}_{\rm{T}}\left( {t - {\tau _0}} \right)\notag\\
=&\frac{{\tilde M}}{\sqrt T}{e^{j2\pi {\nu _0}\left( {t - {\tau _0}} \right)}}\sum\limits_{n =  - \infty }^\infty  {{e^{j2\pi \frac{{\tilde M}}{T}\left( {t - {\tau _0}} \right)}}} {\mathsf {FW}}_{\rm{T}}\left( {t - {\tau _0} - nT} \right){\rm{sinc}}\left( {\frac{{\tilde M}}{T}\left( {t - {\tau _0} - nT} \right)} \right){\mathsf {TW}}_{\rm{T}}\left( {t - {\tau _0}} \right),
\label{DD_basis_truncated_T_with_sinc}
\end{align}
and
\begin{align}
\tilde \Phi _{\rm{F}}^{{\tau _0},{\nu _0}}\left( f \right) =& \tilde NT\left\{ {\Phi _{\rm{F}}^{{\tau _0},{\nu _0}}\left( f \right){\mathsf {FW}}_{\rm{F}}\left( {f - {\nu _0}} \right)} \right\} * \left( {{e^{j2\pi f{\tau _0}}}{e^{ - j\pi \tilde NfT}}{\mathsf {TW}}_{\rm{F}}\left( f \right){\rm{sinc}}\left( {\tilde NTf} \right)} \right)\notag\\
=& \tilde N\sqrt T {e^{ - j\pi f{\tau _0}}}\sum\limits_{m =  - \infty }^\infty  {{e^{ - j\pi \tilde N\left( {f - {\nu _0}} \right)T}}{\mathsf {TW}}_{\rm{F}}\left( {f - {\nu _0}} -\frac{m}{T}\right)} {\rm{sinc}}\left( {\tilde NT\left( {f - {\nu _0} - \frac{m}{T}} \right)} \right){\mathsf {FW}}_{\rm{F}}\left( {\frac{m}{T}} \right).
\label{DD_basis_truncated_F_with_sinc}
\end{align}
Based on~\eqref{DD_basis_truncated_T_with_sinc} and~\eqref{DD_basis_truncated_F_with_sinc}, we observe that both the time and frequency domain basis still follows the pulsone structure, but, instead of the periodic summation of delta functions in~\eqref{time_basis_localized} and~\eqref{freq_basis_localized}, the basis functions are constituted by the periodic summation of the product of the window function and sinc pulse.
\begin{figure*}
\label{TF_basis_function_figures}
\centering
\subfigure[Time domain basis function.]{
\begin{minipage}[t]{0.5\textwidth}
\centering
\includegraphics[scale=0.5]{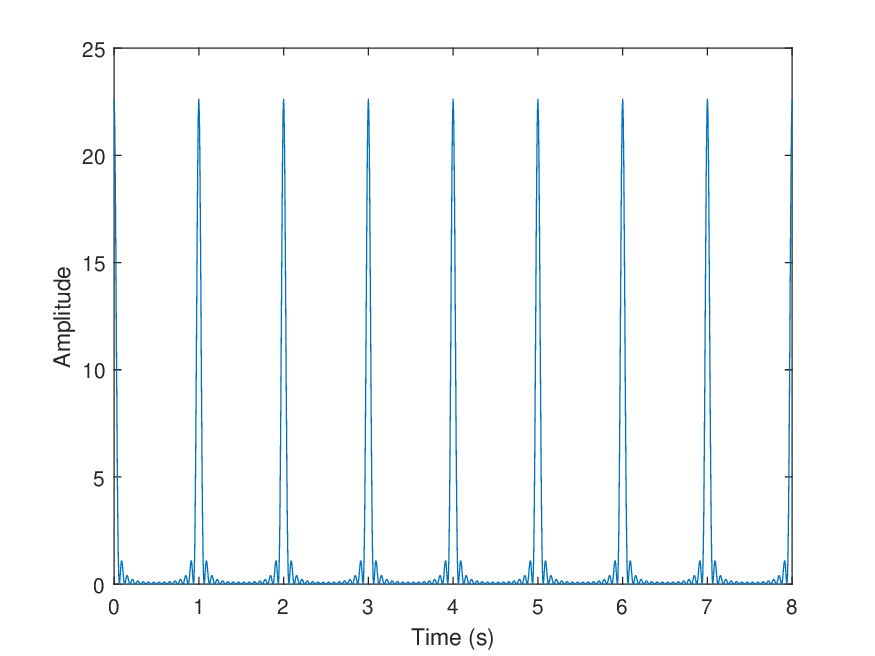}
\label{Time_basis_function_rect_rect}
\end{minipage}%
}%
\subfigure[Frequency domain basis function.]{
\begin{minipage}[t]{0.5\textwidth}
\centering
\includegraphics[scale=0.5]{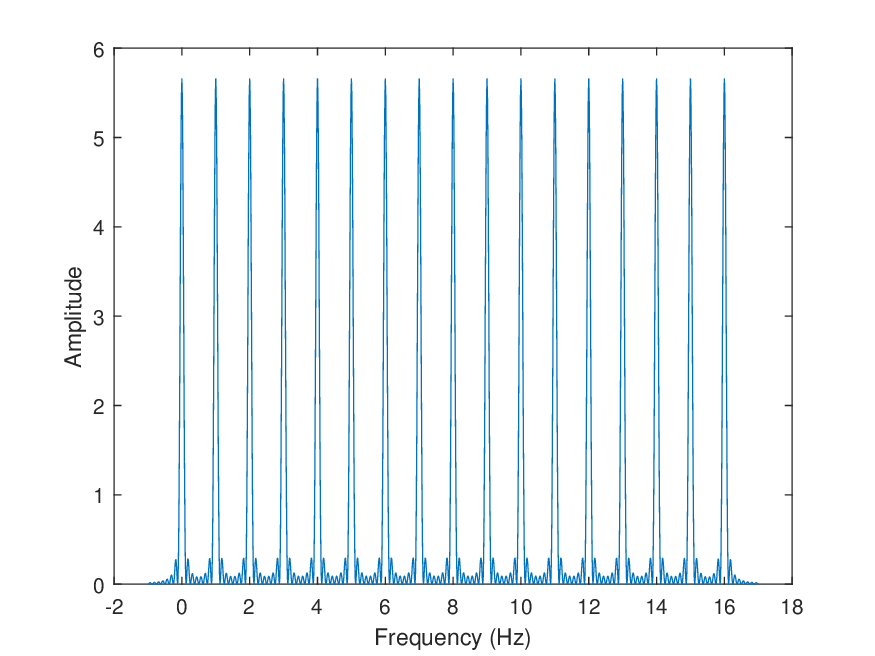}
\label{Freq_basis_function_only_delay}
\end{minipage}%
}%
\caption{Truncated basis functions in time and frequency domains using rectangular windows.}
\end{figure*}
The time and frequency domain basis functions with rectangular windows are shown in Fig.~\ref{Time_basis_function_rect_rect} and Fig.~\ref{Freq_basis_function_only_delay}, where $M=16$, $N=8$, and $T=1$. As shown in the figure, the basis functions in both time and frequency domains remain the pulsone structure, where the `` local pulses'' in time and frequency are sufficiently narrow and are separated by $T$ and $1/T$, respectively. In fact, this is not unexpected as the width of these pulses are roughly determined by the inverse of the windows' bandwidth and time duration~\cite{MohammedBITSpart1,Mohammed2023part2}. We shall refer to this property as the \emph{TF separability property} of the basis function, which holds sufficiently in the asymptotical regime, i.e., sufficiently large $\tilde M$ and $\tilde N$.
Furthermore, we note that the frequency domain signal has slight excessive bandwidth due to the time domain windowing, which is commonly referred to as the OOB emission. Note that the impact of the OOB emission is determined by the underlying time domain window.
\begin{figure*}
\label{DD_basis_function_figures}
\centering
\subfigure[Basis function with 2D rectangular windowing.]{
\begin{minipage}[t]{0.5\textwidth}
\centering
\includegraphics[scale=0.5]{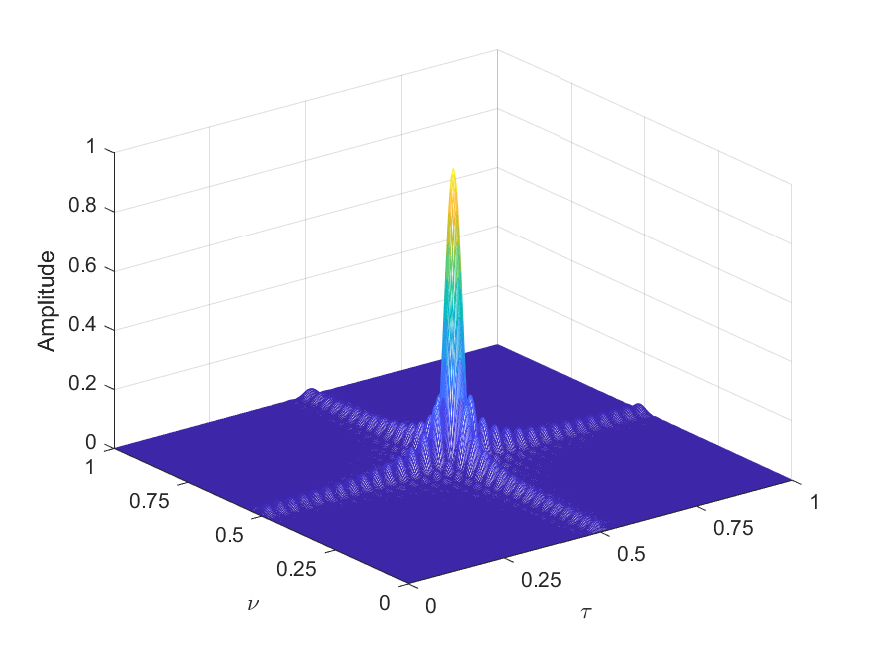}
\label{DD_basis_function_rect_rect}
\end{minipage}%
}%
\subfigure[Basis function with 2D RRC windowing.]{
\begin{minipage}[t]{0.5\textwidth}
\centering
\includegraphics[scale=0.5]{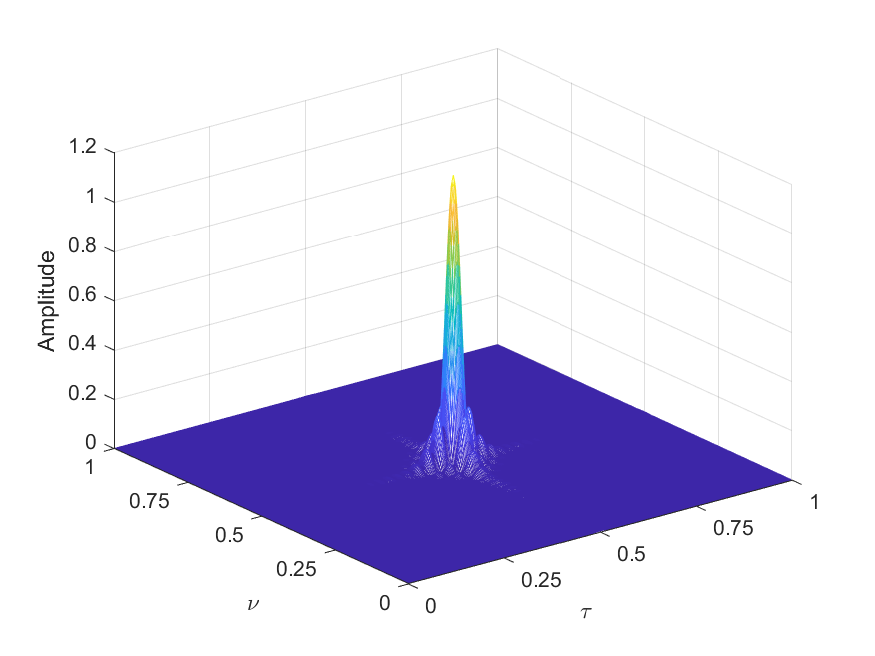}
\label{DD_basis_function_RRC_RRC}
\end{minipage}%
}%
\quad
\subfigure[Delay response.]{
\begin{minipage}[t]{0.5\textwidth}
\centering
\includegraphics[scale=0.5]{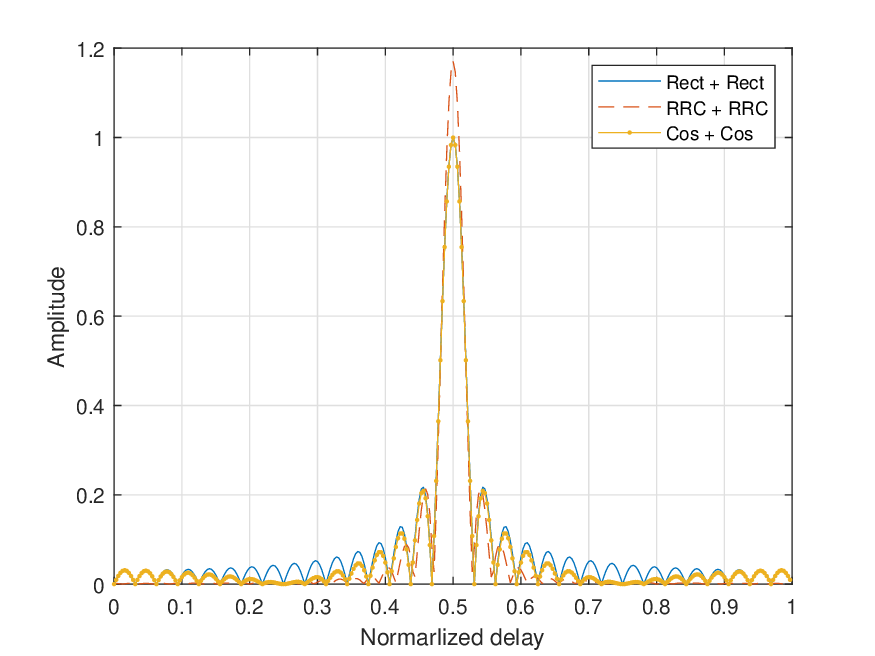}
\label{DD_basis_function_only_delay}
\end{minipage}%
}%
\subfigure[Doppler response.]{
\begin{minipage}[t]{0.5\textwidth}
\centering
\includegraphics[scale=0.5]{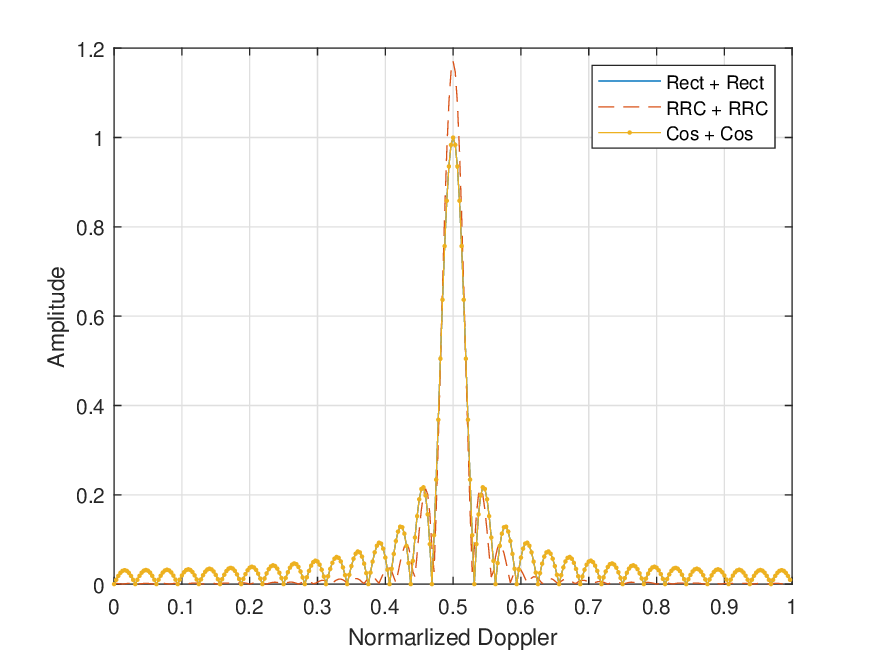}
\label{DD_basis_function_only_Doppler}
\end{minipage}%
}%
\centering
\caption{Truncated DD domain basis functions using different windows.}
\end{figure*}

We demonstrate the truncated DD domain basis functions with different windows in Fig.~\ref{DD_basis_function_rect_rect} to Fig.~\ref{DD_basis_function_only_Doppler}, including the rectangular window (termed ``Rect + Rect''), the root-raised cosine (RRC) window{\footnote{Strictly speaking, we here adopt the Fourier transform of RRC pulses for windowing, i.e., a smoothed rectangular window with excessive bandwidth/time duration.}} with a roll-off factor $0.3$ (termed ``RRC + RRC''), and cosine window (termed ``Cos + Cos''), which is obtained by truncating the continuous cosine signal $\cos \left(t\right)$ from  $t \in \left[ {0,\tilde NT} \right]$ in time and $f \in \left[ {0,\frac{{\tilde M}}{T}} \right]$ in frequency, respectively.
Specifically, we consider $M=N=32$ and $T=1$, and we intentionally consider the DD domain basis function located in the middle of the fundamental rectangle for a better illustration. For both ``Rect + Rect'' and ``Cos + Cos'' cases, we have $\tilde M=M$ and $\tilde N=N$, while for the ``RRC + RRC'' case, $\tilde M$ and $\tilde N$ are slightly larger than $M$ and $N$ to account for the excessive bandwidth/time duration. From Fig.~\ref{DD_basis_function_rect_rect}, we observe that using rectangular windows for basis function truncation results in a sufficiently localized pulse in the DD domain, while suffering from slight power leakage in both the delay and Doppler dimensions. Furthermore, for RRC windows with excessive time and frequency resources, we can see that the truncated basis function is still sufficiently localized but enjoys a much less power leakage in both the delay and Doppler dimensions, as shown in Fig.~\ref{DD_basis_function_RRC_RRC}. This is thanks to the quick decay property of RRC pulses.
The delay and Doppler responses of truncated DD domain basis functions with the three windows are presented in Fig.~\ref{DD_basis_function_only_delay} and Fig.~\ref{DD_basis_function_only_Doppler}. From these two figures, we notice that both ``Rect + Rect'' and ``Cos + Cos' cases share roughly the same delay and Doppler responses, except for the fact that the ``Cos + Cos'' case shows slightly less power leakage around $\tau=0.25$ and $\tau=0.75$. This observation is in line with our derivation in~\eqref{DD_basis_truncated_DD_periodic_window}, where the actual shape of the time domain window only affect the delay domain response slightly. On the other hand, we notice that the ``RRC + RRC'' case indeed enjoys a quick decay in both delay and Doppler dimensions. However, it may not enjoy the perfect orthogonality as ``Rect + Rect'' and ``Cos + Cos'' cases, as some non-zero values appear at integer times of delay and Doppler resolution.

\subsection{The Ambiguity Function of Truncated DD Domain Basis Functions}
In this subsection, we will focus on the ambiguity function of truncated DD domain basis functions. We have demonstrated some properties of truncated DD domain basis functions using some specific windows. Here, we will continue our discussions by highlighting their connections to the ambiguity function.
Notice that after TF-consistent filtering and windowing, the ambiguity function of $\tilde \Phi _{{\rm{DD}}}^{0,0}\left( {\tau ,\nu } \right)$ sufficiently characterizes the DD domain matched-filtering output. Therefore, we are motivated to only consider the ambiguity function of $\tilde \Phi _{{\rm{DD}}}^{0,0}\left( {\tau ,\nu } \right)$, which is defined as ${A_{\tilde \Phi }}\left( {\tau ,\nu } \right)$. In particular, the following theorem characterizes the connections between ${A_{\tilde \Phi }}\left( {\tau ,\nu } \right)$ and the adopted windows.

\textbf{Theorem~3} (\emph{Ambiguity Function of Truncated Basis Function}):
For DD domain basis functions of the form~\eqref{DD_basis_localized} that are frequency and time truncated as~\eqref{DD_basis_truncated_DD}, its ambiguity function satisfies
\begin{align}
{A_{\tilde \Phi }}\left( {\tau ,\nu } \right)=\sum\limits_{n =  - \infty }^\infty  {\sum\limits_{m =  - \infty }^\infty  {{A_{{\mathsf {FW}}}}} \left( {\tau  - nT,\frac{m}{T}} \right)} {A_{\mathsf {TW}}}\left( {\tau ,\nu  - \frac{m}{T}} \right).
\label{AF_trunction}
\end{align}

\textbf{Proof}: The proof is given in Appendix~B. \hfill $\blacksquare$

Theorem~3 states that the ambiguity function of truncated DD domain basis functions can be represented by infinite summations of the ambiguity function of the adopted windows in both time and frequency.
From Theorem~3, various truncated DD domain basis functions with desired ambiguity function can be designed by carefully selecting the windows in time and frequency.
For completeness, we shall also highlight the calculation of ${A_{\tilde \Phi }}\left( {\tau ,\nu } \right)$ from the DD domain by making use of the TF-consistency condition. Recalling~\eqref{DD_MF_basis_functions}, and considering~\eqref{DD_basis_truncated_DD_periodic_window_der1}, after some mathematical manipulations, we arrive at
\begin{align}
{A_{\tilde \Phi }}\left( {\tau_1 ,\nu_1 } \right) &= \int_0^T {\int_0^{\frac{1}{T}} {\tilde \Phi _{{\rm{DD}}}^{0,0}\left( {\tau ,\nu } \right){{\left[ {\tilde \Phi _{{\rm{DD}}}^{{\tau _1},{\nu _1}}\left( {\tau ,\nu } \right)} \right]}^*}} } {\rm{d}}\tau {\rm{d}}\nu \label{AF_trunction_DD_der1}\\
& =\frac{{{{\left| {{\mathsf {FW}}_{\rm{F}}\left( 0 \right)} \right|}^2}}}{T}\sum\limits_{k = 0}^{\tilde N - 1} {{e^{ - j2\pi k{\nu _1}T}}} \sum\limits_{l = 0}^{\tilde M - 1} {\sum\limits_{l' = 0}^{\tilde M - 1} {\int_0^T {{e^{j2\pi l\frac{\tau }{T}}}{e^{ - j2\pi l'\frac{{\tau  - {\tau _1}}}{T}}}{e^{ - j2\pi {\nu _1}\left( {\tau  - {\tau _1}} \right)}}} {\mathsf {TW}}_{\rm{T}}\left( \tau  \right){\mathsf {TW}}_{\rm{T}}^*\left( {\tau  - {\tau _1}} \right){\rm{d}}\tau } }.
\label{AF_trunction_DD}
\end{align}
Based on~\eqref{AF_trunction_DD}, we notice that the Doppler orthogonality can be achieved by general periodic windows in time. However, the delay orthogonality depends on the shape of the adopted time domain window. More precisely, we consider delay and Doppler at integer times of the resolutions, i.e., ${\tau _1} = {l_1}\frac{T}{{\tilde M}}$ and ${\nu _1} = \frac{{{k_1}}}{{\tilde NT}}$ with $ - \tilde M \le {l_1} \le \tilde M$ and $ - \tilde N \le {k_1} \le \tilde N$, which leads to the following theorem.

\textbf{Theorem~4} (\emph{DD Orthogonality}):
Rectangular windows achieve the DD orthogonality, while general periodic windows achieve the DD orthogonality approximately for sufficiently large $\tilde M$ and $\tilde N$. Specifically, we have
\begin{align}
{A_{\tilde \Phi }}\left( {\tau_1 ,\nu_1 } \right) ={\tilde M}{\tilde N}{\left| {{\mathsf {FW}}_{\rm{F}}\left( 0 \right)} \right|^2}{\left| {{\mathsf {TW}}_{\rm{T}}\left( 0 \right)} \right|^2}\delta \left[ {{l_1}} \right]\delta \left[ {{k_1}} \right],
\end{align}
for rectangular windows, and
\begin{align}
{A_{\tilde \Phi }}\left( {\tau_1 ,\nu_1 } \right) \approx \tilde M\tilde N\frac{{{{\left| {{\mathsf {FW}}_{\rm{F}}\left( 0 \right)} \right|}^2}}}{T}\int_0^T {{\mathsf {TW}}_{\rm{T}}\left( \tau  \right){\mathsf {TW}}_{\rm{T}}^*\left( {\tau  - {\tau _1}} \right)} {\rm{d}}\tau \delta \left[ {{l_1}} \right]\delta \left[ {{k_1}} \right],
\end{align}
for general periodic windows.

\textbf{Proof}: The proof is given in Appendix~C. \hfill $\blacksquare$

Theorem~4 essentially states that truncating DD domain basis functions with general periodic windows will result in sufficient orthogonality in both delay and Doppler with respect to $\frac{T}{{\tilde M}}$ and $\frac{1}{{\tilde NT}}$, where the actual shape of the periodic windows do not matter very much as discussed in the previous subsection. Furthermore, using the methodology adopted in Appendix~C, we can show that the ambiguity function also enjoys a semi-periodic response, i.e., the ambiguity function shall have similar responses at $\left( {\tau ,\nu } \right)$ and $\left( {\tau  + nT,\nu  + \frac{m}{T}} \right)$, for any integer $m$ and $n$, given a sufficient time duration and bandwidth of the truncated DD domain basis function{\footnote{In fact, the absolute value of the ambiguity function slightly decreases with larger $n$ and $m$. This is because the the bandwidth and time duration of the transmitted signal are limited.}}. This observation is in line with the quasi-periodicity of the Zak transform. In fact, this is
commonly referred to as the \emph{delay and Doppler ambiguities} in radar theory. However, when the underlying channel is underspread, i.e., the channel satisfies the crystallization condition, the delay spread and Doppler spread are no longer than $T$ and ${\frac{1}{T}}$ as indicated by~\eqref{crystallization}. Such ambiguities do not affect the sensing or communication performance, because the orthogonality is roughly preserved. Furthermore, notice that RRC windows are approximately periodic within their supports for small roll-off factors. Therefore, the DD orthogonality can also be approximately achieved by RRC windows with a smaller roll-off factor.
In fact, the above discussions align well with the response of the truncated DD domain basis function in the fundamental rectangle discussed in the previous subsection. Particularly, we may use either RRC windows or periodic windows interchangeably for achieving DD Nyquist signaling in practice. As a matter of fact, a DD domain signaling of using both RRC window and
rectangular window appears in~\cite{Lin2022ODDM,Hai2022DDOP}.

\textbf{Remark~2}: In fact, the localization in~\eqref{AF_fully_localized} and the orthogonality suggested by Theorem~4 are well-aligned with the intuitions of delay and Doppler by considering the time and frequency partition under critical sampling. Intuitively, the delay and Doppler implies how significant the signal changes in frequency and in time.
Clearly, for periodic signals in time and frequency, their Doppler and delay responses will be fully localized, as will their ambiguity functions. However, for periodic signals with truncation, their delay and Doppler responses will not be fully localized, because the signal periodicity is broken due to the truncation.
Consequently, their Zak transforms will be sufficiently concentrated depending on the duration of the truncation window in time and frequency, following a ``sinc-like'' pattern. This is because the truncation in time and frequency can be viewed as the multiplication of time and frequency rectangular windows, and the ``sinc-like'' pattern appears naturally as the result of the Zak transform to a rectangular window.

\begin{figure*}
\label{AF_figures}
\centering
\subfigure[Zero-Doppler cuts.]{
\begin{minipage}[t]{0.333\textwidth}
\centering
\includegraphics[scale=0.4]{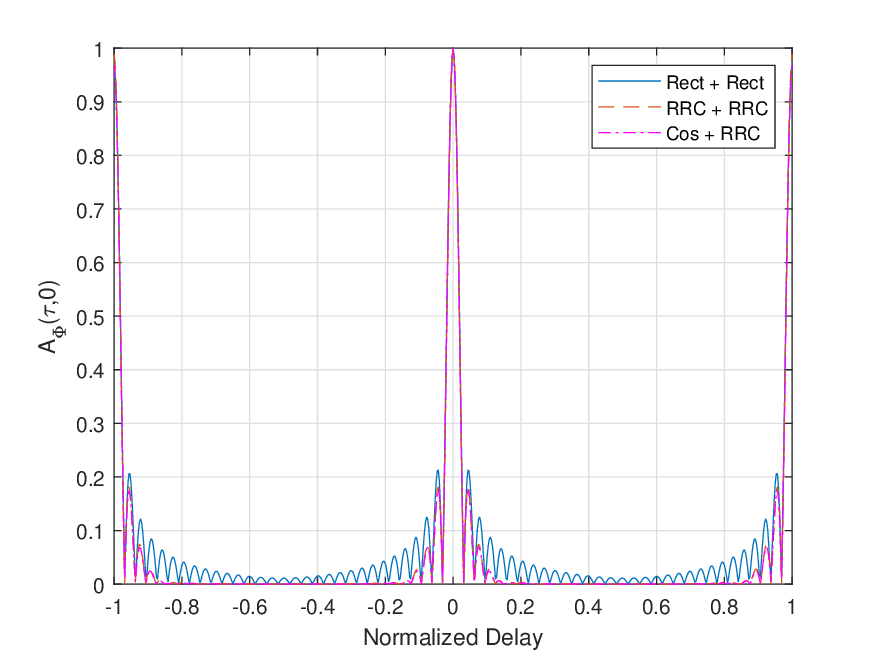}
\label{zero_Doppler_cut}
\end{minipage}%
}%
\subfigure[Zero-delay cuts.]{
\begin{minipage}[t]{0.333\textwidth}
\centering
\includegraphics[scale=0.4]{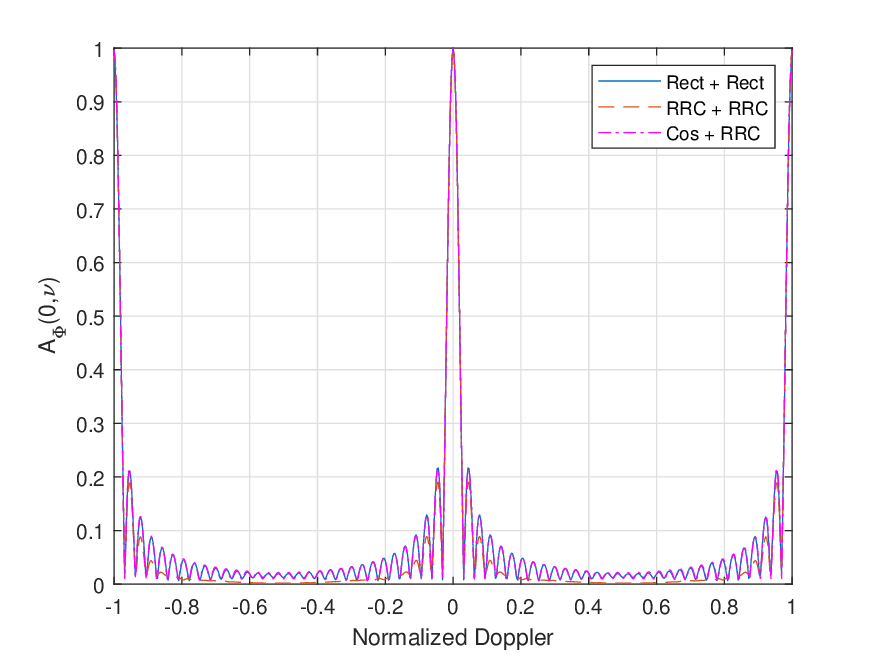}
\label{zero_delay_cut}
\end{minipage}%
}%
\subfigure[Ambiguity function.]{
\begin{minipage}[t]{0.333\textwidth}
\centering
\includegraphics[scale=0.4]{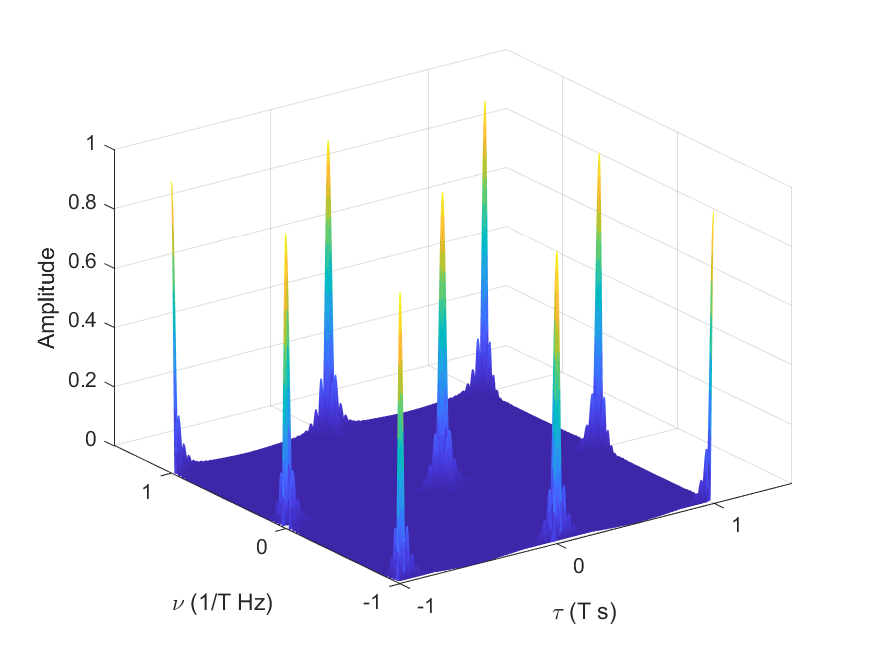}
\label{AF_whole}
\end{minipage}%
}%
\centering
\caption{Truncated DD domain basis functions using different windows.}
\end{figure*}

We demonstrate the DD orthogonality using various time and frequency windows in Fig.~\ref{zero_Doppler_cut} to Fig.~\ref{AF_whole}, where we consider $M=N=32$, and three different windows, namely the rectangular window (termed ``Rect + Rect''), RRC windows with roll-off factor $\beta=0.3$ (termed ``RRC + RRC''), and Cosine window in time and RRC window in frequency (termed as ``Cos + RRC'').
We show the zero-Doppler cuts (ambiguity function with zero Doppler) and the zero-delay cuts (ambiguity function with zero delay) of the three cases in Fig.~\ref{zero_Doppler_cut}, and Fig.~\ref{zero_delay_cut}, where it is observed that all the three cases can achieve the sufficient delay orthogonality. Furthermore, we observe that both (``RRC + RRC'') and (``Cos + RRC'') cases have an almost zero response for normalized delay around $-\frac{T}{2}$ and $\frac{T}{2}$. This is due to the quick decay of RRC pulses at the cost of the excess bandwidth. As a result, we can observe a ``spike-like'' ambiguity function, which may also be of interest for radar sensing.
We also present the plot of ambiguity function for the (``RRC + RRC'') case in Fig.~\ref{AF_whole}, where we observe that the ambiguity function does have a response that is quasi-periodic along both delay and Doppler, while sufficiently localized within the fundamental rectangle.
\section{Practical Pulse Shaping for Delay-Doppler Communications}
In this section, we will discuss the practical pulse shaping for general DD communications and we will also highlight the input-output relation for DD communications with practical pulse shapes.

Recall the TF-consistent filtering and windowing discussed in the previous section. We notice that in order to implement this, multiple time domain and frequency domain windows shall be used to truncate the DD domain basis function.
Particularly, according to~\eqref{DD_modulated_signal} and~\eqref{DD_basis_truncated_T}, the transmitted DD communication signal with roughly limited time and frequency resources can be written by
\begin{align}
{s_{{\rm{T}}}}\left( t\right) =& \sum\limits_{l = 0}^{M - 1} {\sum\limits_{k = 0}^{N - 1} {{x_{{\rm{DD}}}}\left[ {l,k} \right]} } \tilde \Phi _{\rm{T}}^{{\tau _l},{\nu _k}}\left( t \right)\notag\\
=& \sum\limits_{l = 0}^{M - 1} {\sum\limits_{k = 0}^{N - 1} {{x_{{\rm{DD}}}}\left[ {l,k} \right]} } \left\{ {\left[ {{e^{j2\pi {\nu _k}\left( {t - {\tau _l}} \right)}}\Phi _{\rm{T}}^{{\tau _l},{\nu _k}}\left( {t - {\tau _l}} \right)} \right] * \left[ {{\mathsf {FW}}_{\rm{T}}\left( t \right){e^{j2\pi {\nu _k}t}}} \right]} \right\}{\mathsf {TW}}_{\rm{T}}\left( {t - {\tau _l}} \right)\notag\\
= &\sqrt{T} \sum\limits_{l = 0}^{M - 1} {\sum\limits_{k = 0}^{N - 1} {{x_{{\rm{DD}}}}\left[ {l,k} \right]} } {e^{j2\pi {\nu _k}\left( {t - {\tau _l}} \right)}}\sum\limits_{n =  - \infty }^\infty  {{\mathsf {FW}}_{\rm{T}}\left( {t - {\tau _l} - nT} \right)} {\mathsf {TW}}_{\rm{T}}\left( {t - {\tau _l}} \right),\label{DD_com_signal_expand1}
\end{align}
where we assume that both ${{\mathsf {FW}}_{\rm{T}}\left( t \right)}$ and ${{\mathsf {TW}}_{\rm{T}}\left( t \right)}$ have unit power.
According to~\eqref{DD_com_signal_expand1}, it is possible to design the DD pulse shaping using a filter bank structure, which includes $MN$ filters that are shifted in time and frequency with respect to the delay and Doppler offsets associated to ${{x_{{\rm{DD}}}}\left[ {l,k} \right]}$.
Such an implementation is straightforward but may not be practical due to the high hardware complexity required for realizing the filter bank. Therefore, we are motivated to consider a simplified and more practical implementation by applying a \emph{sufficiently narrow} time domain pulse ${\mathsf {FW}}_{\rm{T}}\left( t \right)$. In this case,~\eqref{DD_com_signal_expand1} can be shown to converge to
\begin{align}
{s_{{\rm{T}}}}\left( t\right) \simeq \sqrt{T} \sum\limits_{l = 0}^{M - 1} {\sum\limits_{k = 0}^{N - 1} {{x_{{\rm{DD}}}}\left[ {l,k} \right]} } \sum\limits_{n = -\infty}^{\infty}  {{e^{j2\pi n{\nu _k}T}}{\mathsf {FW}}_{\rm{T}}\left( {t - {\tau _l} - nT} \right)} {\mathsf {TW}}_{\rm{T}}\left( {t} \right).
\label{DD_com_signal_expand2}
\end{align}
The convergence of~\eqref{DD_com_signal_expand2} holds for practical signals, such as RRC signals, when $M$ is large, which can be explained by the TF separability property discussed previously.
Furthermore, by substituting ${\tau _l} = \frac{l}{M}T$ and ${\nu _k} = \frac{k}{{NT}}$ into~\eqref{DD_com_signal_expand2}, we obtain
\begin{align}
{s_{\rm{T}}}\left( t \right) \simeq &\sqrt T \sum\limits_{l = 0}^{M - 1} {\sum\limits_{n = -\infty }^{\infty} {\sum\limits_{k = 0}^{N - 1} {{x_{{\rm{DD}}}}\left[ {l,k} \right]{e^{j2\pi n\frac{k}{N}}}} } } {\mathsf {FW}}_{\rm{T}}\left( {t - \frac{l}{M}T - nT} \right){\mathsf {TW}}_{\rm{T}}\left( t \right)\notag\\
=&\sqrt {NT} \sum\limits_{l = 0}^{M - 1} {\sum\limits_{n = -\infty}^{ \infty} {{\tilde x_{\rm{T}}}\left[ {l + nM} \right]} } {\mathsf {FW}}_{\rm{T}}\left( {t - \frac{l}{M}T - nT} \right){\mathsf {TW}}_{\rm{T}}\left( t \right),
\label{DD_com_signal_expand3}
\end{align}
where
\begin{align}
{\tilde x_{\rm{T}}}\left[ {l + nM} \right] \buildrel \Delta \over = \frac{1}{{\sqrt N }}\sum\limits_{k = 0}^{N - 1} {{x_{{\rm{DD}}}}\left[ {l,k} \right]{e^{j2\pi n\frac{k}{N}}}} ,
\label{IDZT}
\end{align}
is the \emph{inverse discrete Zak transform} (IDZT)~\cite{bolcskei1997discrete} of ${{\bf{X}}_{{\rm{DD}}}}$, and the constant $\sqrt {NT}$ is a normalization factor that roughly agrees with time duration of ${s_{\rm{T}}}\left(t\right)$ in order to maintain the average symbol energy.
Notice that~\eqref{DD_com_signal_expand3} involves an infinite summation of with respect to $n$. Let us define ${{\bf{x}}_{\rm{T}}} \buildrel \Delta \over = {\left[ {{\tilde x_{\rm{T}}}\left[ 0 \right],{\tilde x_{\rm{T}}}\left[ 1 \right],...,{\tilde x_{\rm{T}}}\left[ {MN - 1} \right]} \right]^{\rm{T}}}$ as a length-$MN$ vector, and we shall highlight that ${{\bf{\tilde x}}_{\rm{T}}}$ is a periodically extended version of ${{\bf{x}}_{\rm{T}}}$.
Furthermore, notice that ${\mathsf {TW}}_{\rm{T}}\left( t \right)$ has a time duration roughly $NT$.
Therefore, it is natural to approximate ${{\bf{\tilde x}}_{\rm{T}}}$ by ${{\bf{x}}_{\rm{T}}}$ in practice.
In fact, such an approximation is commonly adopted in the OTFS literature, where ${{\bf x}_{\rm{T}}}$ is often time domain symbol vector for OTFS transmission.
As suggested by~\eqref{DD_com_signal_expand3}, we can achieve DD pulse shaping by filtering ${{\bf x}_{\rm{T}}}$ in a way similar to that of the single-carrier transmission. However, this must be done with care because the underlying wireless channel may introduce additional delay and Doppler shifts.
Therefore, as a common method, we propose to append a sufficiently long cyclic prefix (CP) at the beginning of the frame. Note that the insertion of CP effectively transforms the linear convolution of the time domain channel to the circular convolution. As a result, the received signal can be viewed as a periodized version of the ${{\bf x}_{\rm{T}}}$ shifted by channel delay and Doppler, whose DZT aligns with ${{\bf X}_{\rm{DD}}}$~\cite{Bolcskei1994Gabor}. In fact, this CP structure is commonly referred to as the ``reduced-CP'' structure in the OTFS literature~\cite{Raviteja2019practical}.

\begin{figure}
  \centering
  \includegraphics[width=0.5\textwidth]{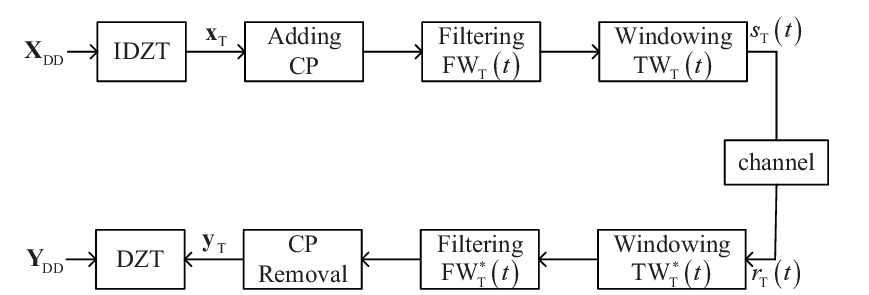}
  \caption{The diagram of the proposed DD Nyquist communication.}
  \label{DD_pulse_shaping}
  \centering
\end{figure}

\begin{figure}
  \centering
  \includegraphics[width=0.5\textwidth]{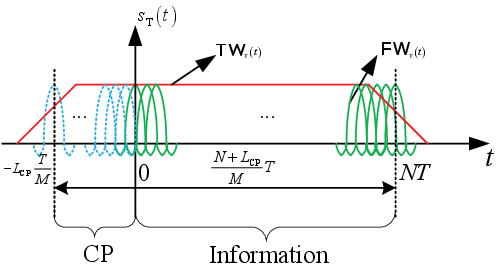}
  \caption{A brief diagram of the time domain transmitted signal with adopted windows, where the CP part and the information part are marked by dashed and solid curves, respectively. Here, $L_{\rm CP}$ denotes the length of CP.}
  \label{DD_pulse_shaping_visualize}
  \centering
\end{figure}

Given the discussions above, we shall consider the DD Nyquist pulse shaping structure as shown in Fig.~\ref{DD_pulse_shaping}. In Fig.~\ref{DD_pulse_shaping}, the DD domain symbol matrix ${\bf X}_{\rm DD}$ is first passed to the IDZT module, yielding ${\bf x}_{\rm T}$ of length-$MN$.
After appending a CP with duration longer than the maximum path delay, the resultant vector is then convoluted with ${\mathsf {FW}}_{\rm{T}}\left( t \right)$ and followed by windowing based on ${\mathsf {TW}}_{\rm{T}}\left( t \right)$, obtaining the time domain transmitted signal ${s_{\rm{T}}}\left( t \right)$. Particularly, we require the adopted time domain window is long enough to cover the CP part as well, as shown in Fig.~\ref{DD_pulse_shaping_visualize}{\footnote{Note that this application does not disobey the approximation of ${\bf \tilde x}_{\rm T}$ by ${\bf x}_{\rm T}$, because the CP preserves the periodicity in time.}}. In the figure, we adopt a time domain window with excessive duration for transmission, where we demonstrate the interaction between time and frequency domain windows. We use dashed and solid curves to mark the CP part and the information part of the signal and $L_{\rm CP}$ here denotes the CP length.

At the receiver side, the time domain received signal ${r_{{\rm{T}}}}\left( t \right)$ is first windowed by ${\mathsf {TW}}_{\rm T}^{*}\left( t \right)$ and then filtered by ${\mathsf {FW}}_{\rm T}^{*}\left( t \right)$, where the connection between ${s_{{\rm{T}}}}\left( t \right)$ and ${r_{{\rm{T}}}}\left( t \right)$ is given by~\eqref{TD_twisted_conv_delay_first_Doppler_second}. After removing the CP, the resultant time domain received symbol vector ${\bf y}_{\rm T}$ is converted to the DD domain via the DZT.
Specifically, we have
\begin{align}
{y_{\rm{T}}}\left[ {l' + n'M} \right] =& \int_{ - \infty }^\infty  {{r_{\rm{T}}}} \left( t \right)\sqrt {NT}{\mathsf {TW}}_{\rm{T}}^*\left( t \right){\mathsf {FW}}_{\rm{T}}^*\left( {t - \frac{{l'}}{M}T - n'T} \right){\rm{d}}t\notag\\
=& {NT} \sum\limits_{p = 1}^P {{h_p}\sum\limits_{l = 0}^{M - 1} {\sum\limits_{n =  - \infty }^\infty  {{x_{\rm{T}}}\left[ {l + nM} \right]} } } \notag\\
&\int_{ - \infty }^\infty  {{e^{j2\pi {{\tilde \nu }_p}\left( {t\! -\! {{\tilde \tau }_p}} \right)}}} {\mathsf {FW}}_{\rm{T}}\left( {t\! - \!\frac{l}{M}T \!-\! nT \!-\! {{\tilde \tau }_p}} \right){\mathsf {TW}}_{\rm{T}}\left( {t\! - \!{\tilde\tau _p}} \right){\mathsf {TW}}_{\rm{T}}^*\left( t \right){\mathsf {FW}}_{\rm{T}}^*\left( {t\! -\! \frac{{l'}}{M}T \!-\! n'T} \right){\rm{d}}t.
\label{Received_MF_Time}
\end{align}
By performing DZT to ${{\bf y}_{\rm{T}}}$, the DD domain received symbol matrix can be obtained, i.e.,
\begin{align}
{y_{{\rm{DD}}}}\left[ {l,k} \right] = \frac{1}{{\sqrt N }}\sum\limits_{n = 0}^{N - 1} {{y_{\rm{T}}}\left[ {l + nM} \right]} {e^{ - j2\pi n\frac{k}{N}}}.
\label{Received_MF_DD}
\end{align}

\subsection{Input-Output Relation for DD Communications with Practical Pulse Shaping}
We now discuss the input-output relation of DD communications with practical pulse shaping.
According to~\eqref{Received_MF_Time}, let us define the effective time domain channel coefficient between the $k$-th transmit symbol and the $k'$-th receive symbol by
\begin{align}
g_{k,k'}^{\left( {\tau ,\nu } \right)} \buildrel \Delta \over = {NT}\int_{ - \infty }^\infty  {{e^{j2\pi \nu \left( {t - \tau } \right)}}} {\mathsf {FW}}_{\rm{T}}\left( {t - \frac{k}{M}T - \tau } \right){\mathsf {TW}}_{\rm{T}}\left( {t - \tau } \right){\mathsf {TW}}_{\rm{T}}^*\left( t \right){\mathsf {FW}}_{\rm{T}}^*\left( {t - \frac{{k'}}{M}T} \right){\rm{d}}t,
\label{time_channel_coef}
\end{align}
where $\left( {\tau ,\nu } \right)$ is a pair of arbitrary delay and Doppler shifts that satisfies the \emph{underspread} channel condition, i.e., $ \tau  \in \left[ {0,T} \right)$, and $\nu  \in \left[ {0,\frac{1}{T}} \right)$. Furthermore, let $L$ be the maximum length of intersymbol interference (ISI), i.e.,
$g_{k,k'}^{\left( {\tau ,\nu } \right)} \approx 0 $ for any $ \tau  \in \left[ {0,T} \right),\nu  \in \left[ {0,\frac{1}{T}} \right)$ and $\left| {k' - k} \right| > L$.
Thus, we obtain
\begin{align}
{{\bf{y}}_{\rm{T}}} = \sum\limits_{p = 1}^P {{\bf{H}}_{\rm{T}}^{\left( p \right)}} {{\bf{x}}_{\rm{T}}} + {{\bf{n}}_{\rm{T}}},
\label{time_effective_io}
\end{align}
where ${{\bf{H}}_{\rm{T}}^{\left( p \right)}}$ is the effective time domain channel matrix of size $MN \times MN$ for the $p$-th resolvable path, given by
\begin{align}
\small
{\bf{H}}_{\rm{T}}^{\left( p \right)} = h_p\left[ {\begin{array}{*{20}{c}}
{g_{0,0}^{\left( {{\tau _p},{\nu _p}} \right)}}& \cdots &{g_{L,0}^{\left( {{\tau _p},{\nu _p}} \right)}}&0& \cdots &0&{g_{ - L,0}^{\left( {{\tau _p},{\nu _p}} \right)}}& \cdots &{g_{ - 1,0}^{\left( {{\tau _p},{\nu _p}} \right)}}\\
 \vdots & \ddots &{}& \ddots & \ddots &{}& \ddots & \ddots & \vdots \\
 \vdots &{}& \ddots &{}&{}&0& \cdots &0&{g_{ - 1,L - 1}^{\left( {{\tau _p},{\nu _p}} \right)}}\\
{g_{0,L}^{\left( {{\tau _p},{\nu _p}} \right)}}& \cdots & \cdots &{g_{L,L}^{\left( {{\tau _p},{\nu _p}} \right)}}& \cdots &{g_{2L,L}^{\left( {{\tau _p},{\nu _p}} \right)}}&0& \cdots &0\\
0&{}&{}&{}&{}&{}&{}&{}&{}\\
{}& \ddots &{}& \ddots &{}& \ddots &{}& \ddots &{}\\
 \vdots &{}&{}&{}&{}&{}&{}&{}& \vdots \\
{}& \ddots &{}& \ddots &{}&{}&{}& \ddots &{}\\
0&{}& \cdots &{}&0&{g_{MN - L - 1,MN - 1}^{\left( {{\tau _p},{\nu _p}} \right)}}& \cdots & \cdots &{g_{MN - 1,MN - 1}^{\left( {{\tau _p},{\nu _p}} \right)}}
\end{array}} \right].
\label{time_effective_H_p}
\end{align}
We observe that ${{\bf{H}}_{\rm{T}}^{\left( p \right)}}$ has a banded structure with small portion of non-zero elements placed on the top-right corner due to the appended CP.
Based on~\eqref{time_effective_H_p}, the effective time domain channel matrix that considers the effect of $P$ paths can be written as ${{\bf{H}}_{\rm{T}}} \buildrel \Delta \over = \sum\nolimits_{p = 1}^P {{\bf{H}}_{\rm{T}}^{\left( p \right)}} $.
Furthermore, notice that the both IDZT and DZT can be described using matrix forms~\cite{li2021cross}. We can then derive the DD domain input-output relation, such as
\begin{align}
{{\bf{y}}_{\rm{DD}}} = {\bf H}_{\rm DD} {{\bf{x}}_{\rm{DD}}} + {{\bf{n}}_{\rm{DD}}},
\label{DD_effective_io}
\end{align}
where ${{\bf{H}}_{{\rm{DD}}}} \buildrel \Delta \over = \left( {{\bf{F}}_N \otimes {{\bf{I}}_M}} \right){{\bf{H}}_{\rm{T}}}\left( {{{\bf{F}}_N^{\rm{H}}} \otimes {{\bf{I}}_M}} \right)$ is the effective DD domain channel matrix and ${{\bf{n}}_{\rm{DD}}}$ is the effective DD domain AWGN vector with the variance of the AWGN samples being $N_0$.

\subsection{Asymptotical DD Domain Input-Output Relation}
In this subsection, we study the simplification of the input-output relation discussed in the previous subsection, by focusing on sufficiently large $M$ and $N$. In such a case, the convergence of time domain pulse shaping in~\eqref{DD_com_signal_expand2} holds generally. Furthermore, following the same reasoning of the transmitter part, we have
\begin{align}
{y_{{\rm{DD}}}}\left[ {l,k} \right] \simeq \int_{ - \infty }^\infty  {{r_{\rm{T}}}\left( t \right)} {\left[ {\tilde \Phi _{\rm{T}}^{{\tau _l},{\nu _k}}\left( t \right)} \right]^*}{\rm{d}}t = \int_0^T {\int_0^{\frac{1}{T}} {{r_{{\rm{DD}}}}\left( {\tau ,\nu } \right)} } {\left[ {\tilde \Phi _{{\rm{DD}}}^{{\tau _l},{\nu _k}}\left( {\tau ,\nu } \right)} \right]^*}{\rm{d}}\tau {\rm{d}}\nu .
\label{TD_convergence_to_DD}
\end{align}
Notice that
\begin{align}
{r_{{\rm{DD}}}}\left( {\tau ,\nu } \right) \simeq& \sum\limits_{p = 1}^P {{h_p}{s_{{\rm{DD}}}}\left( {\tau  - {\tau _p},\nu  - {\nu _p}} \right){e^{j2\pi {\tilde\nu _p}\left( {\tau  - {\tilde\tau _p}} \right)}}}+{n_{{\rm{DD}}}}\left( {\tau ,\nu } \right)\label{DD_com_signal_received0}\\
=& \sum\limits_{p = 1}^P {{h_p}\sum\limits_{l = 0}^{M - 1} {\sum\limits_{k = 0}^{N - 1} {{x_{{\rm{DD}}}}\left[ {l,k} \right]{e^{j2\pi {\tilde\nu _p}\left( {\tau  - {\tilde\tau _p}} \right)}}} } \tilde \Phi _{\rm{DD}}^{{\tau _l},{\nu _k}}\left( {\tau  - {\tilde\tau _p},\nu  - {\tilde\nu _p}} \right)} +{n_{{\rm{DD}}}}\left( {\tau ,\nu } \right),
\label{DD_com_signal_received}
\end{align}
where~\eqref{DD_com_signal_received0} is derived by substituting~\eqref{DD_channel} into~\eqref{DD_twisted_conv_delay_first_Doppler_second}. Then,~\eqref{TD_convergence_to_DD} can be further simplified by
\begin{align}
{y_{{\rm{DD}}}}\left[ {l,k} \right] = \sum\limits_{p = 1}^P {{h_p}\sum\limits_{l' = 0}^{M - 1} {\sum\limits_{k' = 0}^{N - 1} {{x_{{\rm{DD}}}}\left[ {l',k'} \right]\int_0^T {\int_0^{\frac{1}{T}} {{e^{j2\pi {{\tilde \nu }_p}\left( {\tau  - {{\tilde \tau }_p}} \right)}}} } } } \tilde \Phi _{{\rm{DD}}}^{{\tau _{l'}},{\nu _{k'}}}\left( {\tau  - {{\tilde \tau }_p},\nu  - {{\tilde \nu }_p}} \right)} {\left[ {\tilde \Phi _{{\rm{DD}}}^{{\tau _l},{\nu _k}}\left( {\tau ,\nu } \right)} \right]^*}{\rm{d}}\nu {\rm{d}}\tau .
\label{DD_com_signal_MF}
\end{align}
To further characterize~\eqref{DD_com_signal_MF}, let us consider the following corollary.

\textbf{Corollary~2} (\emph{Symbol-wise DD Domain Input-Output Relation}):
Let ${\tilde \Phi _{{\rm{DD}}}^{{\tau _l},{\nu _k}}\left( {\tau ,\nu } \right)}$ and ${\tilde \Phi _{{\rm{DD}}}^{{\tau _{l'}},{\nu _{k'}}}\left( {\tau ,\nu } \right)}$ be the DD domain basis functions associated to the ${{x_{{\rm{DD}}}}\left[ {l,k} \right]}$ and ${{x_{{\rm{DD}}}}\left[ {l',k'} \right]}$, respectively. Furthermore, considering an arbitrary pair of delay and Doppler shifts $\left( {\tilde \tau ,\tilde \nu } \right)$, we have
\begin{align}
&\int_0^T {\int_0^{\frac{1}{T}} {{e^{j2\pi \tilde \nu \left( {\tau  - \tilde \tau } \right)}}\tilde \Phi _{{\rm{DD}}}^{{\tau _{l'}},{\nu _{k'}}}\left( {\tau  - \tilde \tau ,\nu  - \tilde \nu } \right){{\left[ {\tilde \Phi _{{\rm{DD}}}^{{\tau _l},{\nu _k}}\left( {\tau ,\nu } \right)} \right]}^*}} } {\rm{d}}\nu {\rm{d}}\tau  \notag\\
=& {e^{j2\pi \tilde \nu \left( {{\tau _l} - \tilde \tau } \right)}}{e^{j2\pi {\nu _{k'}}\left( {{\tau _l} - \tilde \tau  - {\tau _{l'}}} \right)}}{A_{\tilde \Phi }}\left( {{\tau _l} - \tilde \tau  - {\tau _{l'}},{\nu _k} - \tilde \nu  - {\nu _{k'}}} \right) \label{DD_symbol_wise_io}.
\end{align}

\textbf{Proof}: The proof is given in Appendix~D. \hfill $\blacksquare$

By substituting~\eqref{DD_symbol_wise_io} into~\eqref{DD_com_signal_MF}, we arrive the symbol-wise DD domain input-output relation with large $M$ and $N$ by
\begin{align}
{y_{{\rm{DD}}}}\left[ {l,k} \right] = \sum\limits_{p = 1}^P {{h_p}\sum\limits_{l' = 0}^{M - 1} {\sum\limits_{k' = 0}^{N - 1} {{e^{j2\pi {{\tilde \nu }_p}\left( {{\tau _l} - {{\tilde \tau }_p}} \right)}}{e^{j2\pi {\nu _{k'}}\left( {{\tau _l} - {{\tilde \tau }_p} - {\tau _{l'}}} \right)}}{x_{{\rm{DD}}}}\left[ {l',k'} \right]} } } {A_{\tilde \Phi }}\left( {{\tau _l} - {{\tilde \tau }_p} - {\tau _{l'}},{\nu _k} - {{\tilde \nu }_p} - {\nu _{k'}}} \right).
\label{DD_symbol_wise_io_general}
\end{align}
Notice that~\eqref{DD_symbol_wise_io_general} holds for general pulses/windows without any constraint on the DD orthogonality, where the ambiguity function can be calculated based on~\eqref{AF_trunction}.
To shed the light on practical implementation, we now demonstrate the DD domain input-output relation using special pulses/windows under two specific channel conditions.

\textbf{Example~1} (\emph{Rectangular Windows}): In the case where both power normalized  ${{\mathsf {FW}}_{\rm{F}}\left( f \right)}$ and ${{\mathsf {TW}}_{\rm{T}}\left( t \right)}$ are rectangular windows with bandwidth $\frac{M}{T}$ and time duration $NT$, i.e., ${\mathsf {FW}}_{\rm{F}}\left( f \right) = \frac{1}{{\sqrt {{T \mathord{\left/
 {\vphantom {T M}} \right.
 \kern-\nulldelimiterspace} M}} }}$ only for $f \in \left[ {0,\frac{M}{T}} \right]$ and ${\mathsf {TW}}_{\rm{T}}\left( t \right) = \frac{1}{{\sqrt {NT} }}$ only for $t \in \left[ {0,NT} \right]$. According to~\eqref{AppendixTh4_der2} in Appendix C,~\eqref{DD_symbol_wise_io_general} can be further derived by{\footnote{Here, we again assume that the values of $M$ and $N$ are sufficiently large.}}
\begin{align}
{y_{{\rm{DD}}}}\left[ {l,k} \right] =& \frac{1}{{MN}}\sum\limits_{p = 1}^P {{h_p}\sum\limits_{l' = 0}^{M - 1} {\sum\limits_{k' = 0}^{N - 1} {{e^{j2\pi {{\tilde \nu }_p}\left( {\frac{l}{M}T - {{\tilde \tau }_p}} \right)}}{e^{j2\pi \frac{{k'}}{{NT}}\left( {\frac{{l - l'}}{M}T - {{\tilde \tau }_p}} \right)}}{x_{{\rm{DD}}}}\left[ {l',k'} \right]} } } \notag\\
&\sum\limits_{n = 0}^{N - 1} {{e^{ - j2\pi n\left( {\frac{{k - k'}}{{NT}} - {{\tilde \nu }_p}} \right)T}}\sum\limits_{m = 0}^{M - 1} {{e^{j2\pi m\frac{{\left( {\frac{{l - l'}}{M}T - {{\tilde \tau }_p}} \right)}}{T}}}} } \sum\limits_{m' = 0}^{M - 1} {{e^{j2\pi \left( {\frac{{m' - m}}{T} - \left( {\frac{{k - k'}}{{NT}} - {{\tilde \nu }_p}} \right)} \right)T}}} {\rm{sinc}}\left( {\frac{{m - m'}}{T} - \left( {\frac{{k - k'}}{{NT}} - {{\tilde \nu }_p}} \right)} \right)
\label{DD_symbol_wise_io_rect}
\end{align}
\textbf{Example~2} (\emph{Rectangular Windows with integer delay and Doppler}): In the case of integer delay and Doppler, i.e., ${{\tilde \tau }_p} = \frac{{{{\tilde l}_p}}}{M}T,{{\tilde \nu }_p} = \frac{{{{\tilde k}_p}}}{{NT}}$, where $0 \le {\tilde l}_p \le M - 1$ and $0 \le {{\tilde k}_p} \le N - 1$ are integers for any
$1 \le p\le P$, known as the \emph{delay and Doppler indices},~\eqref{DD_symbol_wise_io_rect} further reduces to
\begin{align}
{y_{{\rm{DD}}}}\left[ {l,k} \right] = &\sum\limits_{p = 1}^P {{h_p}\sum\limits_{l' = 0}^{M - 1} {\sum\limits_{k' = 0}^{N - 1} {{e^{j2\pi \frac{{{{\tilde k}_p}}}{{MN}}\left( {l - {{\tilde l}_p}} \right)}}{e^{j2\pi \frac{{k'}}{{MN}}\left( {l - l' - {{\tilde l}_p}} \right)}}{x_{{\rm{DD}}}}\left[ {l',k'} \right]} } } \notag\\
&\sum\limits_{m =  - \infty }^\infty  {\delta \left[ {k - k' - {{\tilde k}_p} + mN} \right]\sum\limits_{n =  - \infty }^\infty  {\delta \left[ {l - l' - {{\tilde l}_p} + nM} \right]} }\notag\\
=&\sum\limits_{p = 1}^P {{h_p}{e^{j2\pi \frac{{{{\tilde k}_p}}}{{MN}}\left( {l - {{\tilde l}_p}} \right)}}{\alpha _{l,{{\tilde l}_p},k,{{\tilde k}_p}}}} {x_{{\rm{DD}}}}\left[ {{{\left[ {l - {l_p}} \right]}_M},{{\left[ {k - {k_p}} \right]}_N}} \right],
\label{DD_symbol_wise_io_rect_integer}
\end{align}
where
\begin{align}
{\alpha _{l,{{\tilde l}_p},k,{{\tilde k}_p}}} = \left\{ \begin{array}{l}
1,\quad \quad\quad\quad\quad\quad \! l - {l_p} \ge 0\\
{e^{ - j2\pi \frac{{k - {{\tilde k}_p}}}{N}}},\quad\quad l - {l_p} < 0
\end{array} \right.
\label{DD_symbol_wise_io_rect_phase}
\end{align}
is a phase offset due to the DD domain quasi-periodicity~\cite{LSY_THP}. In fact, the above input-output relation aligns with the matrix form provided in~\cite{Raviteja2019practical,Viterbo2022DDcommunications} using rectangular pulses in the TF domain.

\textbf{Remark~3}: We have shown above the input-output relation using rectangular windows, where we observe a consistent input-output relation with the two-stage OTFS implementation under integer delay and Doppler when $M$ and $N$ are sufficiently large. This is not unexpected, because OFDM signaling with rectangular windows can be implemented in the time domain via \emph{sinc interpolation}. This scheme is commonly known as the discrete multi-tone (DMT) transmission~\cite{Alphan2014survey}.

\section{Numerical Results}
In this section, we present the numerical results on the proposed DD communications in terms of BER and pragmatic capacity. Without loss of generality, we consider $M=16$ and $N=16$, unless specified otherwise and the transmitted symbols are obtained from an energy-normalized QPSK constellation. Furthermore, we assume that the underspread wireless channel has $P=4$ independent resolvable paths, where the delay indices and Doppler indices can have fractional values and are uniformly taking values from $\left[0,l_{\max}\right]$ and $\left[-k_{\max}/2,k_{\max}/2\right]$, respectively, with $l_{\max}=5$ and $k_{\max}=3$. We will compare the pulse shaping using rectangular windows and RRC windows with roll-off
factors $\beta=0.3$ in the frequency domain and $\beta=0.1$ in the time domain.

\begin{figure*}
\label{DD_vs_OFDM_performance}
\centering
\subfigure[BER rate for fractional DD.]{
\begin{minipage}[t]{0.5\textwidth}
\centering
\includegraphics[scale=0.5]{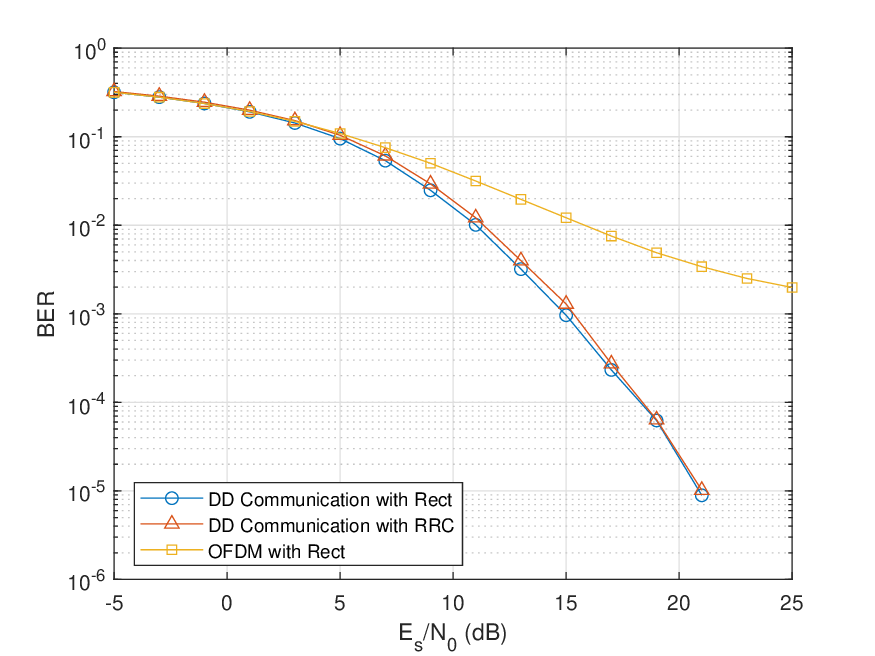}
\label{BER_fractional_DD}
\end{minipage}%
}%
\subfigure[BER rate for integer DD.]{
\begin{minipage}[t]{0.5\textwidth}
\centering
\includegraphics[scale=0.5]{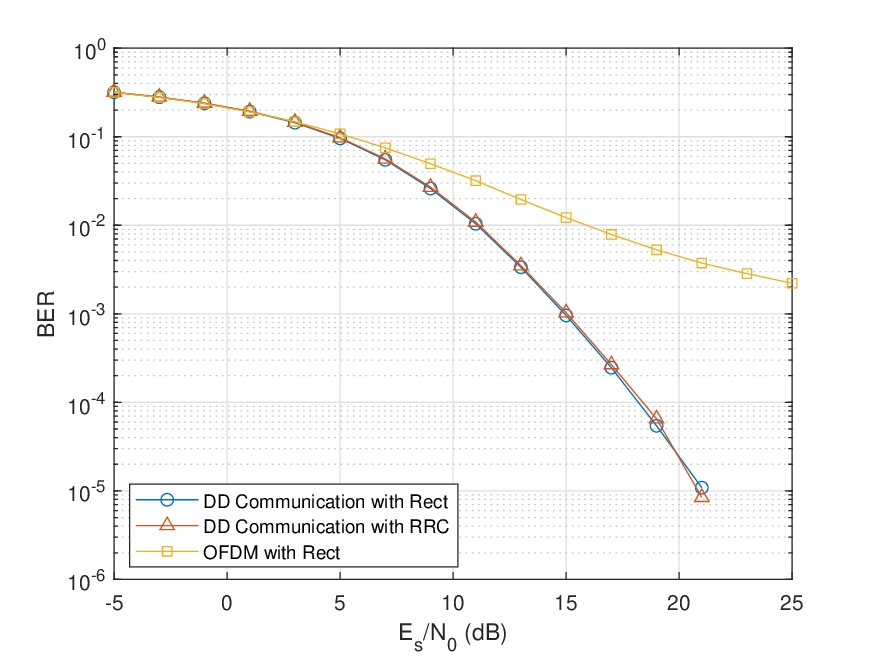}
\label{BER_integer_DD}
\end{minipage}%
}%
\quad
\subfigure[Pragmatic capacity for fractional DD.]{
\begin{minipage}[t]{0.5\textwidth}
\centering
\includegraphics[scale=0.5]{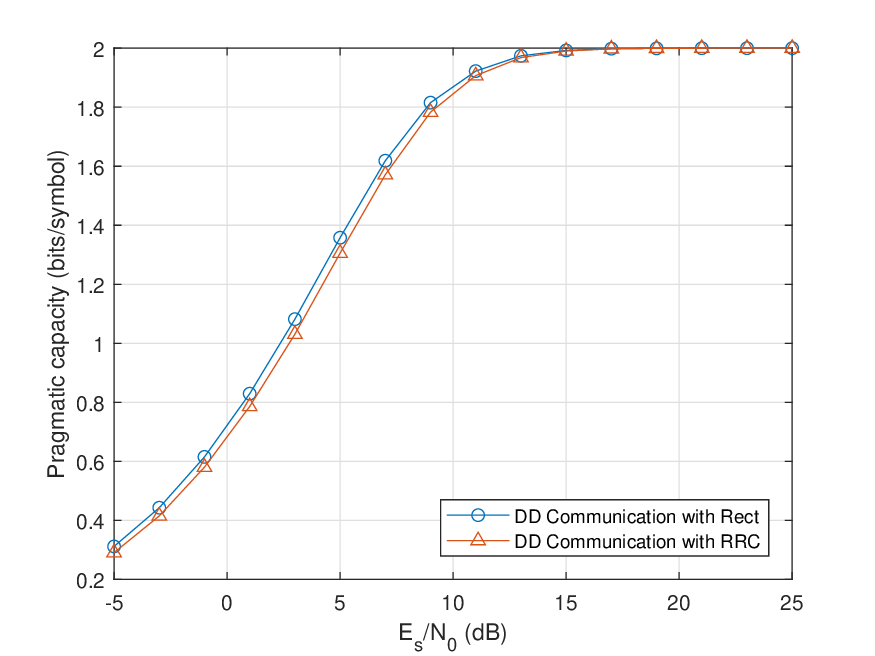}
\label{PC_fractional_DD}
\end{minipage}%
}%
\subfigure[PSD of DD communication signals.]{
\begin{minipage}[t]{0.5\textwidth}
\centering
\includegraphics[scale=0.5]{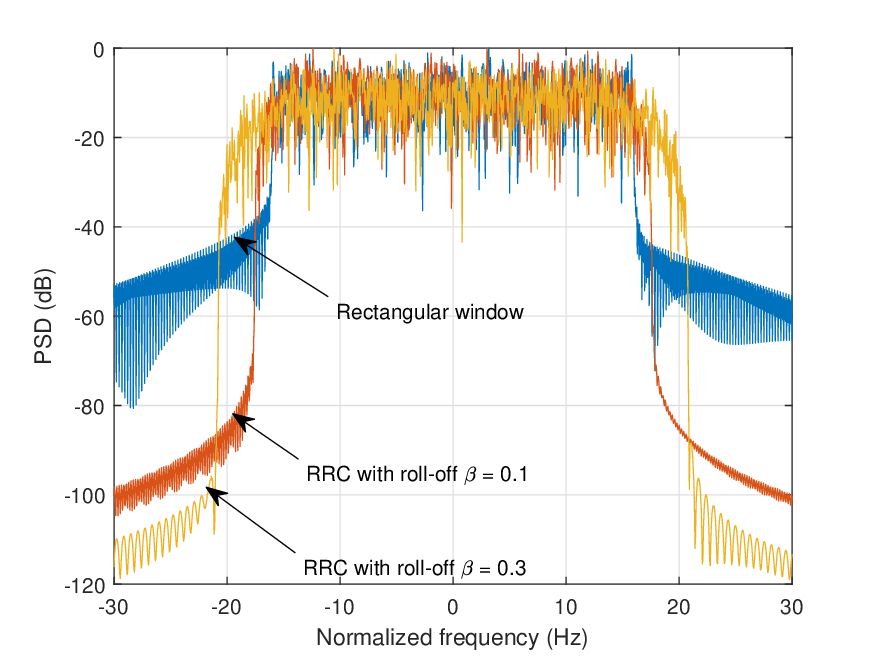}
\label{PSD_MN168_0103}
\end{minipage}%
}%
\centering
\caption{Performance evaluation of DD communications and its comparison to OFDM.}
\end{figure*}

We present the BER performance of the proposed transmission in comparison with OFDM under the same channel condition in Fig.~\ref{BER_fractional_DD} and Fig.~\ref{BER_integer_DD}, where the OFDM is implemented using the DMT structure with sinc pulses (rectangular windows).
To attain a good error performance, we adopt the cross domain iterative detection proposed in~\cite{li2021cross} for both DD communications and OFDM, which is shown to provide a near-optimal performance for OTFS. We notice from Fig.~\ref{BER_fractional_DD} and Fig.~\ref{BER_integer_DD} that the DD communications with both rectangular and RRC windows can obtain roughly the same error performance under both fractional and integer delay and Doppler case, especially in the high SNR regime. This is because the cross domain iterative detection can achieve the near-optimal performance that is dominated by the channel fading rather than the exact delay and Doppler response.
Furthermore, we highlight that the application of time domain RRC windows may result in slight reduction in the effective symbol energy of few transmitted symbols due to the roll-off part as indicated by Fig.~\ref{DD_pulse_shaping_visualize}. However, this power loss is almost has no influence on the error performance when we choose a relatively small roll-off factor for the time domain RRC window, e.g., $\beta=0.1$.
On the other hand, OFDM in both two cases fails to provide good BER results. Particularly, we notice that the BER curves of OFDM have a less steep slope compared to those of DD communications, which indicates that it cannot fully exploit the full channel diversity.
However, we highlight that the above results are obtained from uncoded systems. A more detailed analysis between coded OTFS and OFDM appears in~\cite{Li2020performance}.




The pragmatic capacity performance of the considered schemes under the fractional DD case are presented in Fig.~\ref{PC_fractional_DD}. The pragmatic capacity is characterized by the ``single-letter" mutual information incorporating the effects of modulation, channel, demodulation, and equalization, which can be computed numerically by using Monte Carlo simulations. More details of pragmatic capacity can be found in~\cite{Lorenzo2022fair,Kavcic2003binary}. From the figure, we observe that the pragmatic capacities of DD communications using both rectangular and RRC windows achieve roughly the same performance, despite the fact that using rectangular window can achieve a slight rate improvement in the low-to-mid SNR regime. This observation align well with our BER curves in Fig.~\ref{BER_fractional_DD}.

We finally examine the PSDs of the transmitted signals in Fig.~\ref{PSD_MN168_0103}, which are obtained by taking the squared values of the Fourier transform of~\eqref{DD_com_signal_expand3}. Specifically, we present the PSD corresponding to the rectangular window, and RRC windows with roll-off factors with $\beta=0.1$ and $\beta=0.3$, and we consider $M=16$ and $N=8$ here.
It is not surprising to see that the PSD of transmitted signals using RRC window will exhibit a much lower OOB emission compared to the case of using rectangular windows. Particularly, we observe that the lower OOB emission comes at a price of the excess bandwidth determined by the roll-off factor, and a larger roll-off factor is beneficial for obtaining a lower OOB emission.
This observation verifies the practical advantage of the proposed DD communications.

\section{Conclusion}
In this paper, we discussed the practical implementation of DD communications based on insights from the Zak transform. We firstly presented our basis function construction and then highlighted its features in different domains. We then presented the practical realization of the constructed basis functions based on TF-consistent windowing and filtering, and derived their ambiguity functions, where we verified the sufficient DD orthogonality can be achieved by truncated periodic signals. Finally, we derived the end-to-end system model for the proposed scheme under various shaping pulses and our conclusions were verified by our numerical results.

\appendices

\section{Derivation of~\eqref{AF_fully_localized}}
By substituting~\eqref{DD_basis_localized} into~\eqref{DD_MF_basis_functions},
we have
\begin{align}
&{e^{j2\pi {\nu _2}\left( {{\tau _1} - {\tau _2}} \right)}}{A_\Phi }\left( {{\tau _1} - {\tau _2},{\nu _1} - {\nu _2}} \right)\notag\\
=& \int_0^T {\int_0^{\frac{1}{T}} {{e^{j2\pi {\nu _2}\left( {\tau  - {\tau _2}} \right)}}\Phi _{{\rm{DD}}}^{0,0}\left( {\tau  - {\tau _2},\nu  - {\nu _2}} \right){e^{ - j2\pi {\nu _1}\left( {\tau  - {\tau _1}} \right)}}{{\left( {\Phi _{{\rm{DD}}}^{0,0}\left( {\tau  - {\tau _1},\nu  - {\nu _1}} \right)} \right)}^*}} } {\rm{d}}\nu {\rm{d}}\tau \notag\\
=&\sum\limits_{n =  - \infty }^\infty  {\sum\limits_{m =  - \infty }^\infty  {\int_{ - \infty }^\infty  {\int_{ - \infty }^\infty  {{e^{j2\pi {\nu _2}\left( {\tau  - {\tau _2}} \right)}}{e^{ - j2\pi {\nu _1}\left( {\tau  - {\tau _1}} \right)}}\varphi \left( {\tau  - {\tau _2} - nT,\nu  - {\nu _2} - \frac{m}{T}} \right)} } {\varphi ^*}\left( {\tau  - {\tau _1},\nu  - {\nu _1}} \right)} } {e^{ - j2\pi n\left( {\nu  - {\nu _2}} \right)T}}{\rm{d}}\nu {\rm{d}}\tau ,\label{AppendixB_der1}
\end{align}
where~\eqref{AppendixB_der1} is derived by interchanging the variables and replacing the summation by extending the integral range.
Furthermore, by substituting $\varphi \left( {\tau ,\nu } \right) = \delta \left( \tau  \right)\delta \left( \nu  \right)$ into~\eqref{AppendixB_der1}, we obtain
\begin{align}
{e^{j2\pi {\nu _2}\left( {{\tau _1} - {\tau _2}} \right)}}{A_\Phi }\left( {{\tau _1} - {\tau _2},{\nu _1} - {\nu _2}} \right)= &{e^{j2\pi {\nu _2}\left( {{\tau _1} - {\tau _2}} \right)}}\sum\limits_{n =  - \infty }^\infty  {\sum\limits_{m =  - \infty }^\infty  \delta  } \left( {{\tau _1} - {\tau _2} - nT} \right)\delta \left( {{\nu _1} - {\nu _2} - \frac{m}{T}} \right).\label{AppendixB_der2}
\end{align}
Finally, by deleting the phase term on both sides of~\eqref{AppendixB_der2},~\eqref{AF_fully_localized} can be derived. \hfill $\blacksquare$

\section{Proof of Theorem~3}
Define an intermediate function ${\Phi}_{\rm T}'\left( t \right) \buildrel \Delta \over =  \Phi _{\rm{T}}^{0,0}\left( t \right) * {\mathsf {FW}}_{\rm T}\left( t \right) = \int_{ - \infty }^\infty  {{\mathsf {FW}}_{\rm T}\left( \tau  \right)} \Phi _{\rm{T}}^{0,0}\left( {t - \tau } \right){\mathop{\rm d}\nolimits} \tau $. Then, the ambiguity function of ${\Phi}_{\rm T}'\left( t \right)$ can be calculated by
\begin{align}
{A_{{{\Phi}_{\rm T}'}}}\left( {\tau ,\nu } \right)
=&\int_{ - \infty }^\infty  {\int_{ - \infty }^\infty  {{\mathsf {FW}}_{\rm T}\left( {{\tau _1}} \right)} \Phi _{\rm{T}}^{0,0}\left( {t - {\tau _1}} \right){\mathop{\rm d}\nolimits} {\tau _1}\int_{ - \infty }^\infty  {{\mathsf {FW}}^*_{\rm T}\left( {{\tau _2}} \right)} {{\left[ {\Phi _{\rm{T}}^{0,0}\left( {t - {\tau _2} - \tau } \right)} \right]}^*}{\mathop{\rm d}\nolimits} {\tau _2}} {e^{ - j2\pi \nu \left( {t - \tau } \right)}}{\rm{d}}t\notag\\
=&\int_{ - \infty }^\infty  {\int_{ - \infty }^\infty  {{\mathsf {FW}}_{\rm T}\left( {{\tau _1}} \right)} {\mathsf {FW}}^*_{\rm T}\left( {{\tau _2}} \right){e^{ - j2\pi \nu {\tau _2}}}{A_\Phi }\left( {{\tau _2} - {\tau _1} + \tau ,\nu } \right){\mathop{\rm d}\nolimits} {\tau _1}{\mathop{\rm d}\nolimits} {\tau _2}}\notag\\
=&\sum\limits_{n =  - \infty }^\infty  {\int_{ - \infty }^\infty  {{\mathsf {FW}}_{\rm T}\left( {{\tau _1}} \right)} {\mathsf {FW}}^*_{\rm T}\left( {{\tau _1} + nT - \tau } \right){\mathop{\rm d}\nolimits} {\tau _1}{e^{ - j2\pi \nu \left( {{\tau _1}+nT - \tau } \right)}}\sum\limits_{m =  - \infty }^\infty  {\delta \left( {\nu  - \frac{m}{T}} \right)} }  \notag\\
=&\sum\limits_{n =  - \infty }^\infty  {{A_{{\mathsf {FW}}}}\left( { \tau -nT,\nu} \right)\sum\limits_{m =  - \infty }^\infty  {\delta \left( {\nu  - \frac{m}{T}} \right)} }  .\label{AppendixC_der1}
\end{align}
Notice that $\tilde \Phi _{\rm{F}}^{0,0}\left( f \right) = {\Phi}_{\rm F}'\left( f\right) * {\mathsf {TW}}_{\rm{F}}\left( f \right)=\int_{ - \infty }^\infty  {{\mathsf {TW}}_{\rm{F}}\left( \nu  \right){{\Phi}_{\rm F}'}\left( {f - \nu } \right)} {\rm{d}}\nu $, where ${{\Phi}_{\rm F}'}\left( {f } \right)$ is the Fourier transform of ${{\Phi}_{\rm T}'}\left( t \right)$. The ambiguity function of $\tilde \Phi _{\rm{T}}^{0,0}\left( t \right)$ is then calculated by
\begin{align}
{A_{\tilde \Phi }}\left( {\tau ,\nu } \right)
=&\int_{ - \infty }^\infty  {\int_{ - \infty }^\infty  {{\mathsf {TW}}_{\rm{F}}\left( {{\nu _1}} \right)} {{\Phi '}_{\rm{F}}}\left( {f - {\nu _1}} \right){\mathop{\rm d}\nolimits} {\nu _1}\int_{ - \infty }^\infty  {{\mathsf {TW}}^*_{\rm{F}}\left( {{\nu _2}} \right)} {{\left[ {{{\Phi '}_{\rm{F}}}\left( {f - {\nu _2} - \nu } \right)} \right]}^*}{\mathop{\rm d}\nolimits} {\nu _2}} {e^{j2\pi f\tau }}{\rm{d}}f\notag\\
=&\int_{ - \infty }^\infty  {\int_{ - \infty }^\infty  {{\mathsf {TW}}_{\rm{F}}\left( {{\nu _1}} \right){\mathsf {TW}}_{\rm{F}}^*\left( {{\nu _2}} \right){A_{\Phi _{\rm{T}}'}}\left( {\tau ,{\nu _2} - {\nu _1} + \nu } \right){e^{j2\pi {\nu _1}\tau }}} } {\rm{d}}{\nu _1}{\rm{d}}{\nu _2} \notag\\
=&\sum\limits_{n =  - \infty }^\infty  {\sum\limits_{m =  - \infty }^\infty  {{A_{{\mathsf {FW}}}}\left( {\tau  - nT,\frac{m}{T}} \right)\int_{ - \infty }^\infty  {{\mathsf {TW}}_{\rm{F}}\left( {{\nu _1}} \right)} {\mathsf {TW}}_{\rm{F}}^*\left( {{\nu _1} - \left( {\nu  - \frac{m}{T}} \right)} \right){e^{j2\pi {\nu _1}\tau }}{\rm{d}}{\nu _1}} }  \notag\\
=&\sum\limits_{n =  - \infty }^\infty  {\sum\limits_{m =  - \infty }^\infty  {{A_{{\mathsf {FW}}}}} \left( {\tau  - nT,\frac{m}{T}} \right)} {A_{{\mathsf {TW}}}}\left( {\tau ,\nu  - \frac{m}{T}} \right) .\label{AppendixC_der2}
\end{align}

This completes the proof for Theorem~3.\hfill $\blacksquare$
\section{Proof of Theorem~4}
In the case where the time domain window is a rectangular window{\footnote{Here, we assume that ${\mathsf {TW}}_{\rm{T}}\left( t  \right)$ is centered at $t=0$. Note that the following derivations hold trivially for ${\mathsf {TW}}_{\rm{T}}\left( t  \right)$ centered at other $t = \tilde N /2$, by noticing the Zak transform of time-shifted signal results can be represented by the same DD domain response with an additional phase term, i.e.,~\eqref{Zak_quasi_periodicity_delay}, which does not impact our conclusions.}}, where ${\mathsf {TW}}_{\rm{T}}\left( \tau  \right) = {\mathsf {TW}}_{\rm{T}}\left( {\tau  - {\tau _1}} \right) = {\mathsf {TW}}_{\rm{T}}\left( 0 \right)$,~\eqref{AF_trunction_DD} is further simplified by
\begin{align}
{A_{\tilde \Phi }}\left( {{\tau _1},{\nu _1}} \right) =&
\frac{{{{\left| {{\mathsf {FW}}_{\rm{F}}\left( 0 \right)} \right|}^2}{{\left| {{\mathsf {TW}}_{\rm{T}}\left( 0 \right)} \right|}^2}}}{T}\sum\limits_{k = 0}^{\tilde N - 1} {{e^{ - j2\pi k{\nu _1}T}}} \sum\limits_{l = 0}^{\tilde M - 1} {\sum\limits_{l' = 0}^{\tilde M - 1} {\int_0^T {{e^{j2\pi l\frac{\tau }{T}}}{e^{ - j2\pi l'\frac{{\tau  - {\tau _1}}}{T}}}{e^{ - j2\pi {\nu _1}\left( {\tau  - {\tau _1}} \right)}}} {\rm{d}}\tau } } \notag\\
=& \frac{{{{\left| {{\mathsf {FW}}_{\rm{F}}\left( 0 \right)} \right|}^2}{{\left| {{\mathsf {TW}}_{\rm{T}}\left( 0 \right)} \right|}^2}}}{T}\sum\limits_{k = 0}^{\tilde N - 1} {{e^{ - j2\pi k{\nu _1}T}}} \sum\limits_{l' = 0}^{\tilde M - 1} {{e^{j2\pi l'\frac{{{\tau _1}}}{T}}}\sum\limits_{l = 0}^{\tilde M - 1} {{e^{j2\pi {\nu _1}{\tau _1}}}\underbrace {\int_0^T {{e^{j2\pi \left( {l - l' - {\nu _1}T} \right)\frac{\tau }{T}}}} {\rm{d}}\tau }_{T{e^{j2\pi \left( {\frac{{l - l'}}{T} - {\nu _1}} \right)T}}{\rm{sinc}}\left( {\frac{{l - l'}}{T} - {\nu _1}} \right)}} } \notag\\
=& {\left| {{\mathsf {FW}}_{\rm{F}}\left( 0 \right)} \right|^2}{\left| {{\mathsf {TW}}_{\rm{T}}\left( 0 \right)} \right|^2}{e^{j2\pi {\nu _1}{\tau _1}}}\sum\limits_{k = 0}^{\tilde N - 1} {{e^{ - j2\pi k{\nu _1}T}}} \sum\limits_{l' = 0}^{\tilde M - 1} {{e^{j2\pi l'\frac{{{\tau _1}}}{T}}}\sum\limits_{l = 0}^{\tilde M - 1} {{e^{j2\pi \left( {\frac{{l - l'}}{T} - {\nu _1}} \right)T}}} } {\rm{sinc}}\left( {\frac{{l - l'}}{T} - {\nu _1}} \right).
 \label{AppendixTh4_der2}
\end{align}
Furthermore, when ${\tau _1} = {l_1}\frac{T}{{\tilde M}}$ and ${\nu _1} = \frac{{{k_1}}}{{\tilde NT}}$ with $ - \tilde M \le {l_1} \le \tilde M$ and $ - \tilde N \le {k_1} \le \tilde N$,~\eqref{AppendixTh4_der2} is further simplified by
\begin{align}
{A_{\tilde \Phi }}\left( {{\tau _1},{\nu _1}} \right) =&{\tilde N}{\left| {{\mathsf {FW}}_{\rm{F}}\left( 0 \right)} \right|^2}{\left| {{\mathsf {TW}}_{\rm{T}}\left( 0 \right)} \right|^2}\delta \left[ {{k_1}} \right]\sum\limits_{l' = 0}^{\tilde M - 1} {{e^{j2\pi l'\frac{{{\tau _1}}}{T}}}\sum\limits_{l = 0}^{\tilde M - 1} {{e^{j2\pi \left( {\frac{{l - l'}}{T}} \right)T}}} } \delta \left[ {l - l'} \right]\notag\\
=& {\tilde M}{\tilde N}{\left| {{\mathsf {FW}}_{\rm{F}}\left( 0 \right)} \right|^2}{\left| {{\mathsf {TW}}_{\rm{T}}\left( 0 \right)} \right|^2}\delta \left[ {{l_1}} \right]\delta \left[ {{k_1}} \right].
 \label{AppendixTh4_der3}
\end{align}
On the other hand, in the case when the time domain window is not rectangular, we further write~\eqref{AF_trunction_DD} in the form the frequency domain integral, i.e.,
\begin{align}
{A_{\tilde \Phi }}\left( {{\tau _1},{\nu _1}} \right) =& \frac{{{{\left| {{\mathsf {FW}}_{\rm{F}}\left( 0 \right)} \right|}^2}}}{T}{e^{j2\pi {\nu _1}{\tau _1}}}\sum\limits_{k = 0}^{\tilde N - 1} {{e^{ - j2\pi k{\nu _1}T}}} \sum\limits_{l = 0}^{\tilde M - 1} {\sum\limits_{l' = 0}^{\tilde M - 1} {{e^{j2\pi l'\frac{{{\tau _1}}}{T}}}} }  \notag\\
&\int_{ - \infty }^\infty  {\int_{ - \infty }^\infty  {\int_0^T {{e^{j2\pi \left( {{f_1} - {f_2} - {\nu _1} + \frac{{l - l'}}{T}} \right)\tau }}{\rm{d}}\tau } } } {\mathsf {TW}}_{\rm{F}}\left( {{f_1}} \right){\mathsf {TW}}_{\rm{F}}^*\left( {{f_2}} \right){e^{j2\pi {f_2}{\tau _1}}}{\rm{d}}{f_1}{\rm{d}}{f_2} \notag\\
=&{\left| {{\mathsf {FW}}_{\rm{F}}\left( 0 \right)} \right|^2}{e^{j2\pi {\nu _1}{\tau _1}}}\sum\limits_{k = 0}^{\tilde N - 1} {{e^{ - j2\pi k{\nu _1}T}}} \sum\limits_{l = 0}^{\tilde M - 1} {\sum\limits_{l' = 0}^{\tilde M - 1} {{e^{j2\pi l'\frac{{{\tau _1}}}{T}}}}{\int_{ - \infty }^\infty  {\int_{ - \infty }^\infty  {{\mathsf {TW}}_{\rm{F}}\left( {{f_1}} \right){\mathsf {TW}}_{\rm{F}}^*\left( {{f_2}} \right)} } } }  \notag\\
&{e^{j2\pi {f_2}{\tau _1}}}{e^{j2\pi \left( {{f_1} - {f_2} - {\nu _1} + \frac{{l - l'}}{T}} \right)T}}{\rm{sinc}}\left( {\left( {{f_1} - {f_2} - {\nu _1} + \frac{{l - l'}}{T}} \right)T} \right){\rm{d}}{f_1}{\rm{d}}{f_2} . \label{AppendixTh4_der4}
\end{align}
When ${\tau _1} = {l_1}\frac{T}{{\tilde M}}$ and ${\nu _1} = \frac{{{k_1}}}{{\tilde NT}}$ with $ - \tilde M \le {l_1} \le \tilde M$ and $ - \tilde N \le {k_1} \le \tilde N$,~\eqref{AppendixTh4_der4} is further simplified by
\begin{align}
{A_{\tilde \Phi }}\left( {{\tau _1},{\nu _1}} \right) =& {\tilde N}{\left| {{\mathsf {FW}}_{\rm{F}}\left( 0 \right)} \right|^2}\delta \left[ {{k_1}} \right]\sum\limits_{l = 0}^{\tilde M - 1} {\sum\limits_{l' = 0}^{\tilde M - 1}{{e^{j2\pi l'\frac{{{\tau _1}}}{T}}}} {\int_{ - \infty }^\infty  {\int_{ - \infty }^\infty  {{\mathsf {TW}}_{\rm{F}}\left( {{f_1}} \right){\mathsf {TW}}_{\rm{F}}^*\left( {{f_2}} \right)} } } } \notag\\
&{e^{j2\pi {f_2}{\tau _1}}}{e^{j2\pi \left( {{f_1} - {f_2} + \frac{{l - l'}}{T}} \right)T}}{\rm{sinc}}\left( {\left( {{f_1} - {f_2} + \frac{{l - l'}}{T}} \right)T} \right){\rm{d}}{f_1}{\rm{d}}{f_2} \label{AppendixTh4_der5}
\end{align}
Furthermore, we consider the TF separability property, which states that ${\mathsf {TW}}_{\rm{F}}\left( {{f_1}} \right){\mathsf {TW}}_{\rm{F}}^*\left( {{f_2}} \right) \approx 0$, if $\left| {{f_1} - {f_2}} \right| > 1/T$.
Then,~\eqref{AppendixTh4_der5} can be equivalently calculated based on~\eqref{AF_trunction_DD} by admitting $l=l'$, which yields
\begin{align}
{A_{\tilde \Phi }}\left( {{\tau _1},{\nu _1}} \right) \approx  &\frac{{{{\left| {{\mathsf {FW}}_{\rm{F}}\left( 0 \right)} \right|}^2}}}{T}\tilde N\delta \left[ {{k_1}} \right]\sum\limits_{l = 0}^{\tilde M - 1} {{e^{j2\pi l\frac{{{\tau _1}}}{T}}}\int_0^T {{\mathsf {TW}}_{\rm{T}}\left( \tau  \right){\mathsf {TW}}_{\rm{T}}^*\left( {\tau  - {\tau _1}} \right)} {\rm{d}}\tau } \notag\\
=& \tilde M\tilde N\frac{{{{\left| {{\mathsf {FW}}_{\rm{F}}\left( 0 \right)} \right|}^2}}}{T}\int_0^T {{\mathsf {TW}}_{\rm{T}}\left( \tau  \right){\mathsf {TW}}_{\rm{T}}^*\left( {\tau  - {\tau _1}} \right)} {\rm{d}}\tau \delta \left[ {{l_1}} \right]\delta \left[ {{k_1}} \right].
\label{AppendixTh4_der6}
\end{align}
This completes the proof of Theorem~4.\hfill $\blacksquare$

\section{Proof of Corollary~2}
By considering~\eqref{DD_basis_offset}, we have
\begin{align}
&\int_0^T {\int_0^{\frac{1}{T}} {{e^{j2\pi \tilde \nu \left( {\tau  - \tilde \tau } \right)}}\tilde \Phi _{{\rm{DD}}}^{{\tau _{l'}},{\nu _{k'}}}\left( {\tau  - \tilde \tau ,\nu  - \tilde \nu } \right){{\left[ {\tilde \Phi _{{\rm{DD}}}^{{\tau _l},{\nu _k}}\left( {\tau ,\nu } \right)} \right]}^*}} } {\rm{d}}\nu {\rm{d}}\tau \notag\\
=& \int_0^T {\int_0^{\frac{1}{T}} {{e^{j2\pi \tilde \nu \left( {\tau  - \tilde \tau } \right)}}{e^{j2\pi {\nu _{k'}}\left( {\tau  - \tilde \tau  - {\tau _{l'}}} \right)}}\tilde \Phi _{{\rm{DD}}}^{0,0}\left( {\tau  - \tilde \tau  - {\tau _{l'}},\nu  - \tilde \nu  - {\nu _{k'}}} \right){{\left[ {\tilde \Phi _{{\rm{DD}}}^{{\tau _l},{\nu _k}}\left( {\tau ,\nu } \right)} \right]}^*}} } {\rm{d}}\nu {\rm{d}}\tau \notag\\
=& {e^{j2\pi \tilde \nu {\tau _{l'}}}}\int_0^T {\int_0^{\frac{1}{T}} {{e^{j2\pi \left( {\tilde \nu  + {\nu _{k'}}} \right)\left( {\tau  - \tilde \tau  - {\tau _{l'}}} \right)}}\tilde \Phi _{{\rm{DD}}}^{0,0}\left( {\tau  - \tilde \tau  - {\tau _{l'}},\nu  - \tilde \nu  - {\nu _{k'}}} \right){{\left[ {\tilde \Phi _{{\rm{DD}}}^{{\tau _l},{\nu _k}}\left( {\tau ,\nu } \right)} \right]}^*}} } {\rm{d}}\nu {\rm{d}}\tau \notag\\
=&{e^{j2\pi \tilde \nu \left( {{\tau _l} - \tilde \tau } \right)}}{e^{j2\pi {\nu _{k'}}\left( {{\tau _l} - \tilde \tau  - {\tau _{l'}}} \right)}}{A_{\tilde \Phi }}\left( {{\tau _l} - \tilde \tau  - {\tau _{l'}},{\nu _k} - \tilde \nu  - {\nu _{k'}}} \right),
 \label{AppendixTh5_der1}
\end{align}
where~\eqref{AppendixTh5_der1} can be derived based on~\eqref{DD_basis_function_00_vs_AF}.
This completes the proof for Corollary~2.\hfill $\blacksquare$

\section{Acknowledgement}
The authors would like to express their gratitude to the many OTFS experts for the related discussion at the OTFS 2.0 workshop in IEEE Global Communication Conference 2023.

\ifCLASSOPTIONcaptionsoff
  \newpage
\fi

\bibliographystyle{IEEEtran}
\bibliography{OTFS_references}

\end{document}